\documentclass[reprint,superscriptaddress,aps,prl]{revtex4-2}

\usepackage[english]{babel}
\usepackage{amsmath}
\usepackage{amssymb}
\usepackage{graphicx}
\usepackage[colorlinks=true, allcolors=blue]{hyperref}
\usepackage{cleveref}
\usepackage{calrsfs}
\usepackage{mathtools}

\usepackage{soul,xcolor}
\setstcolor{red}

\newcommand{\lp}{\left(}
\newcommand{\rp}{\right)}
\newcommand{\lbk}{\left[}
\newcommand{\rbk}{\right]}
\newcommand{\nn}{\nonumber}

\renewcommand{\selectlanguage}[1]{}

\begin{document}

\title{Nonequilibrium Transitions in a Template Copying Ensemble}

\author{Arthur Genthon}
\affiliation{Max Planck Institute for the Physics of Complex Systems, 01187 Dresden, Germany}
\author{Carl D. Modes}
\affiliation{Max Planck Institute for Molecular Cell Biology and Genetics, 01307 Dresden, Germany}
\affiliation{Center for Systems Biology Dresden, 01307 Dresden, Germany}
\affiliation{Cluster of Excellence, Physics of Life, TU Dresden, 01307 Dresden, Germany}
\author{Frank Jülicher}
\affiliation{Max Planck Institute for the Physics of Complex Systems, 01187 Dresden, Germany}
\affiliation{Center for Systems Biology Dresden, 01307 Dresden, Germany}
\affiliation{Cluster of Excellence, Physics of Life, TU Dresden, 01307 Dresden, Germany}
\author{Stephan W. Grill}
\affiliation{Max Planck Institute for Molecular Cell Biology and Genetics, 01307 Dresden, Germany}
\affiliation{Center for Systems Biology Dresden, 01307 Dresden, Germany}
\affiliation{Cluster of Excellence, Physics of Life, TU Dresden, 01307 Dresden, Germany}

\begin{abstract}
The fuel-driven process of replication in living systems generates distributions of copied entities with varying degrees of copying accuracy. Here we introduce a thermodynamically consistent ensemble for investigating universal population features of template copying systems. In the context of copolymer copying, coarse-graining over molecular details, we establish a phase diagram of copying accuracy. We discover sharp non-equilibrium transitions between populations of random and accurate copies. Maintaining a population of accurate copies requires a minimum energy expenditure that depends on the configurational entropy of copolymer sequences.
\end{abstract}

\maketitle

{\it Introduction -} 
The ability to replicate is a hallmark of the living world. 
Organisms can replicate themselves, as well as cells, and 
DNA replication, RNA transcription and RNA translation to proteins are examples of polymer template copying \cite{kornberg_dna_1992}. 
Generating a DNA polymer copy with near-identical sequence to the template DNA requires energy consumption \cite{lieberman_enzymatic_1955,yarranton_enzyme-catalyzed_1979}. DNA replication is catalyzed by the molecule DNA polymerase, which progressively moves along the template DNA strand as it generates a polymer copy \cite{lehman_enzymatic_1958}. Generating copies of DNA competes with DNA disassembly, catalyzed for example by DNAses without involvement of a fuel \cite{yang_nucleases_2011}.
Detailed models have been used to discuss the key properties of this copy process, typically focusing on individual copies of a template sequence
\cite{andrieux_nonequilibrium_2008,andrieux_molecular_2009,sartori_kinetic_2013,gaspard_kinetics_2014,sartori_thermodynamics_2015,rao_thermodynamics_2015,ouldridge_fundamental_2017,poulton_nonequilibrium_2019,chiuchiu_error-speed_2019,sahoo_accuracy_2021,poulton_edge-effects_2021,juritz_minimal_2022}. This provided insights onto the fundamental limits and trade-offs associated with template copying. Examples of this include trade-offs (or absence thereof) \cite{bennett_dissipation-error_1979,rao_thermodynamics_2015,sahoo_accuracy_2021,qureshi_information_2024} and correlations \cite{chiuchiu_error-speed_2019} between speed, accuracy and cost of copying ; links between dissipation, elongation and information transmission \cite{andrieux_nonequilibrium_2008,sartori_kinetic_2013}; and definitions of copying efficiency \cite{poulton_nonequilibrium_2019}.

Here we investigate the conditions for establishing whole populations of accurate copies of a copolymer template. 
In order to focus on generic features, we coarse-grain 
molecular details of copying such as sequential steps of initiation \cite{poulton_edge-effects_2021}, polymer elongation \cite{andrieux_nonequilibrium_2008,gaspard_kinetics_2014} and strand separation \cite{poulton_nonequilibrium_2019,juritz_minimal_2022} into a one-step stochastic process. We define what we call a template copying ensemble where a single template in presence of reservoirs of fuel and monomers generates a population of stochastic copies. 
We study the distribution of copying errors as a function of copying specificity and active driving by the fuel.
We establish a phase diagram of copying accuracy for the template copying ensemble, and discover sharp transitions between populations of random and accurate copies in the limit of long polymers. 
Our template copying ensemble allows for a thermodynamic description of non-equilibrium steady-state populations of accurate and random polymer copies.

\begin{figure}
    \centering
    \includegraphics[width=\linewidth]{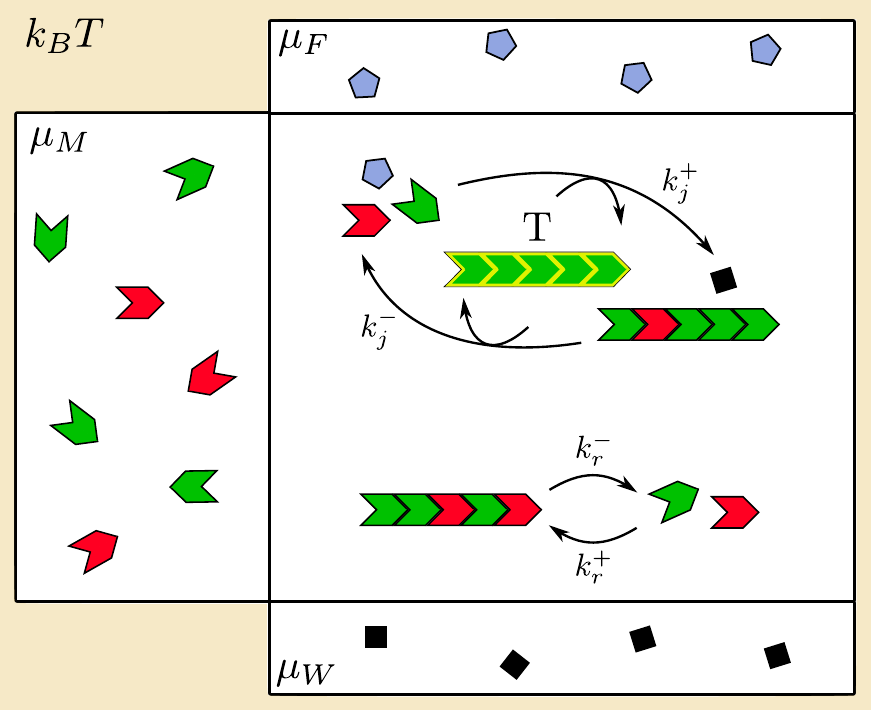}
    \caption{Schematic of the template copying ensemble with two monomer types ($m=2$): green and red. In this example, the template T is composed of green monomers, thus red monomers in sequences S$_j$ are incorrectly copied.}
    \label{fig:ensemble}
\end{figure}

{\it Template copying ensemble -}
We consider a system containing one polymer template sequence T of length $L$ in contact with four reservoirs: a pool of monomers of $m$ different types M$_i$ with $i=1\dots m$ ($m=4$ for DNA and RNA), a bath of fuel molecules F,  a bath of waste molecules W, and a heat bath at constant temperature (we measure  energy in units of the thermal energy, $k_B T=1$).  We call the setting the template copying ensemble. 
Sequences S$_j$ with $j=1\dots m^L$ can be generated by copying the template sequence in a process we refer to as templated assembly. 
The templated assembly process consumes fuel F and generates waste W, to produce copy sequences S$_j$ of the same length as the template without altering the template:
\begin{equation}
    \sum_{i=1}^m n_{ij} \mathrm{M}_i + \mathrm{T} + L \mathrm{F} \ \xrightleftharpoons[k_j^-]{k_j^+} \ \mathrm{S}_j + \mathrm{T} + L \mathrm{W} \,.
\label{rct:copy}
\end{equation}
Here $n_{ij}$ is a stoichiometric coefficient describing the number of monomers of type $i$ in sequence S$_j$, with $\sum_{i=1}^{m} n_{ij} = L$. The templated assembly leading to sequence S$_j$ occurs at a rate $k_{j}^+$ and we consider that template availability is not limiting. Copies are error free if $k_j^+= 0$ for all sequences $j$ that differ from the template, $ \mathrm{S}_j\neq \mathrm{T}$. Copying errors are captured by finite rates $k_j^+> 0$ for these sequences. Microscopic reversibility implies that for each $j$ the reverse pathway, which we call templated disassembly, also exists with rate $k_{j}^-$. However, one may expect  spontaneous disassembly at rate $k_{r}^-$ to be more frequent:
\begin{equation}
    \mathrm{S}_j \ \xrightleftharpoons[k_r^+]{k_r^-} \ \sum_{i=1}^m n_{ij} \mathrm{M}_i \,,
    \label{rct:copy_spont}
\end{equation}
with $k_{r}^+$ denoting a spontaneous assembly rate.

Microreversibility requires that the templated assembly and disassembly rates obey $k_j^+/k_j^- = e^{-(\Delta \mu_r - \Delta \mu_F)L}$,
where $\Delta \mu_r=\epsilon_S/L - \mu_M$ is the per-monomer energy associated with the assembly of a single polymer, and
$\Delta \mu_F  = \mu_F-\mu_W >0$  is the per-monomer Gibbs free energy provided by the fuel. Because $\Delta \mu_r$ is independent of the template, its dependence on sequence S$_j$ cannot be used to generate accurate copies of the template \cite{ouldridge_fundamental_2017}. 
The template behaves as a catalyst and kinetic rates and energy barriers depend on template sequence.  
We therefore choose $\Delta \mu_r$ to be independent of sequence S$_j$. 
We write the rates as $k_j^+ = k_j e^{-(\Delta \mu_r - \Delta \mu_F)L}$, and for the reverse rate $k_j^-=  k_j$. Sequence dependence of the process enters via the kinetic coefficients $k_j$ according to
$k_j=k_0 e^{-a q}$ \cite{SM},
where $q \leq L$ is the number of incorrectly copied monomers (the Hamming distance between T and S$_j$), $k_0$ is a rate prefactor, and parameter $a$ a specificity.  

The spontaneous disassembly and assembly rates also obey $k_r^+/k_r^- = e^{-\Delta \mu_r L}$.
We write for the rate of spontaneous assembly $k_r^+=k_r e^{-\Delta \mu_r L}$ and  for the rate of spontaneous disassembly $k_r^-=k_r$, with a sequence-independent coefficient $k_r$. 

Note that for now the forward rates $k_j^+$ and $k_r^+$ depend on energetics but the backward rates $k_j^-$ and $k_r^-$ do not. 
In this case, the rate of templated disassembly $k_j^-=k_0 e^{-a q}$ vanishes for large $q$ while the rate of spontaneous disassembly $k_r^-$ is constant.
In a later section we will relax this assumption. 

A schematic representation of the system and reservoirs is provided in \cref{fig:ensemble}. Our coarse-grained model describes copying as a one-step process.  Polymers of length different from $L$ could occur as intermediate states but are not considered at the coarse-grained level.

\begin{figure}
    \centering
    \includegraphics[width=\linewidth]{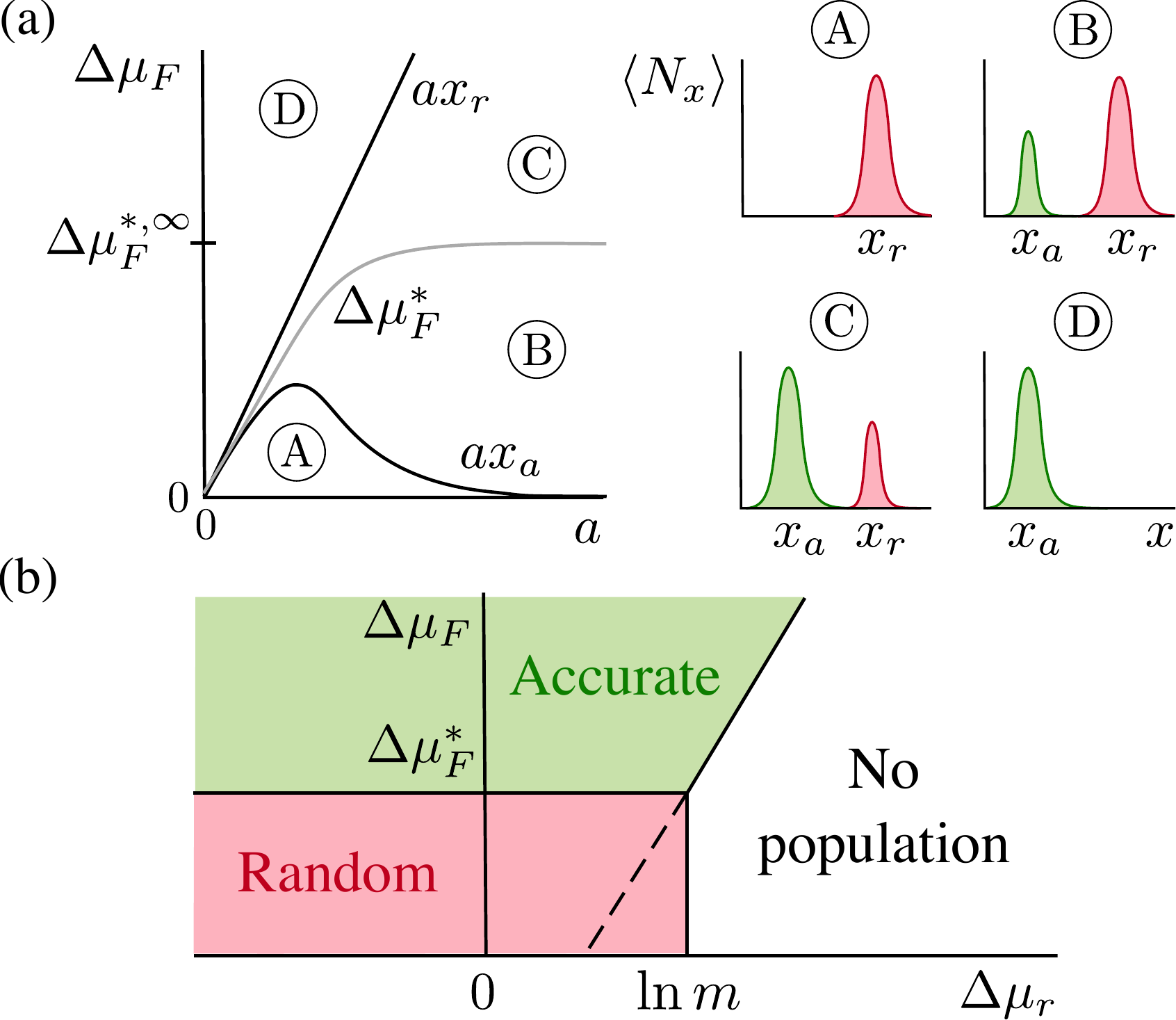}
    \caption{Schematic phase diagrams for (a) finite $L$ and (b) large $L$ with a fixed value of specificity $a$. 
    (a) $\Delta \mu_F^{*,\infty} = \ln m + O(L^{-1})$ is the asymptotic value of $\Delta \mu_F^{*}$ for large specificity $a$.}
    \label{fig:phase_diag}
\end{figure}

{\it Statistics of copying errors -}
We next determine the probability distribution $p(N_S,t)$ to have $N_S$ copies of sequence S at time $t$ which obeys
\begin{align}
	\partial_t p(N_S,t) = & k_a p(N_S-1,t) - (k_a + N_S k_d) p(N_S,t) \nn \\
    & + (N_S+1) k_d p(N_S+1,t) \\
	\partial_t p(0,t) =& -k_a p(0,t) + k_d p(1,t)
    \label{eq:mast_eq}
\end{align}
with the total assembly rate $k_a=k_j^+ + k_r^+$ and the total disassembly rate $k_d=k_j^- + k_r^-$. 
We choose as initial condition $p(N_S,0) = \delta_{N_s}$, yielding a Poisson distribution $p(N_S,t)=\lambda_q^{N_S} e^{-\lambda_q}/N_S!$ for all times, with $ \lambda_q=k_a/k_d (1- \exp{(-k_d t)})$ \cite{SM}. The expected number of copies $\langle N_x \rangle$ with a monomer error fraction $x=q/L$ is
$\langle N_x \rangle =\lambda_{xL} \Omega_{xL}$
with $\Omega_q= {L \choose q} (m-1)^q$ the number of sequences with $q$ wrong monomers. For sufficiently long polymers $L \gg 1$,  $\langle N_x \rangle$  is dominated by either random copies with error fraction $x_r$ or by accurate copies with error fraction $x_a$, or both (see \cref{fig:phase_diag}a). The average fraction of copying errors $\overline{x} =  \sum_x x  \langle N_x \rangle  /\sum_x  \langle N_x \rangle$ is used as a measure of copying accuracy. At first order in $1/L$, the error frequencies $x_i$ with index $i \in \{r,a\}$ are given by
$x_i = x_i^{(0)} - \lp 1-2 x_i^{(0)} \rp/(2L) + o \lp L^{-1} \rp$ with
\begin{align}
    x_r^{(0)} &= \frac{m-1}{m} \\
    \label{eq_xa}
    x_a^{(0)} &= \frac{1}{1 + e^{a}/(m-1)} \, .
\end{align}
When $a \to 0$ the templated copying process becomes nonspecific and $x_a^{(0)} \to x_r^{(0)}$. 
When $a \to + \infty$ the templated copying process becomes precise and $x_a^{(0)} \to 0$. 

The distribution of copying errors $\langle N_x \rangle$ depends on the Gibbs free energy provided by the fuel $\Delta \mu_F$. We next consider the copying errors in steady-state.
If $\Delta \mu_F \leq a x_a$ the number of accurate copies is small, and random sequences dominate. If $\Delta \mu_F \geq a x_r$, the number of random copies is small, and the templated assembly process 
dominates. Under both conditions $\langle N_x \rangle$ is unimodal with a maximum at $x=x_r$ and $x =x_a$, respectively.
These modes correspond to the dominant error fraction of the sequences assembled via the spontaneous and templated processes, respectively. 
If instead $a x_a < \Delta \mu_F < a x_r$, random and accurate copies co-exist, and $\langle N_x \rangle$ is bimodal. In this regime, the number of random copies $\langle N_{x_r} \rangle$ and the number of accurate copies $\langle N_{x_a} \rangle$ are equal when $ \Delta \mu_F = \Delta \mu_F^*$, with
\begin{align}
        \Delta \mu_F^*= &\ln \lp \frac{m}{1+(m-1) e^{-a}} \rp \nn \\
        & + \frac{1}{L} \ln \lp \frac{k_r}{k_0}\sqrt{\frac{x_a^{(0)}(1-x_a^{(0)})}{x_r^{(0)}(1-x_r^{(0)})}}\rp + o\lp\frac{1}{L}\rp \,.
    \label{eq_muf*}
\end{align}
These cases are represented in \cref{fig:phase_diag}a. The dependence of
$\Delta \mu_F^*$  on specificity $a$ is shown in \cref{fig:phase_diag}a in grey.

\begin{figure}
    \centering
    \includegraphics[width=\linewidth]{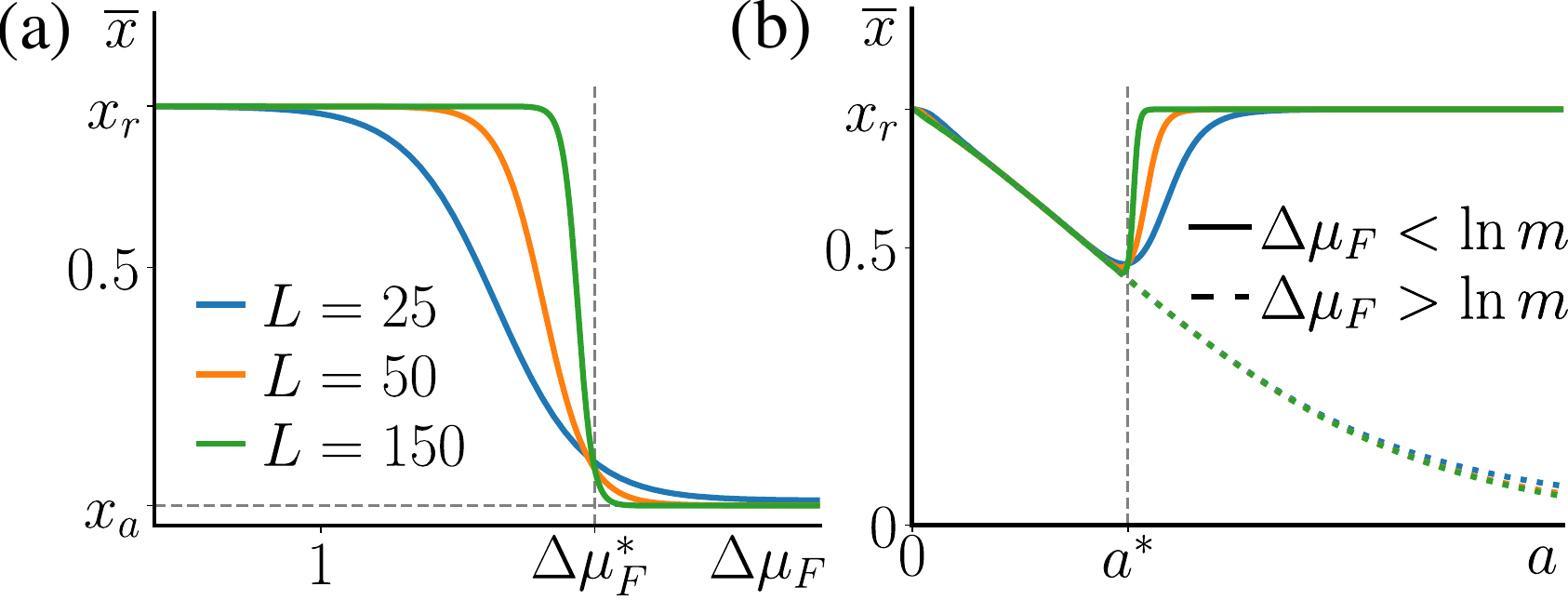}
    \caption{Average error fraction $\overline{x}$ vs (a) energy drive $\Delta \mu_F$ and (b) specificity $a$, for different values of the template length $L$ (colors apply to (a) and (b)). (a,b) $m=4$, $k_0=1$, $k_r=0.1$ and $\Delta \mu_r=0.5$. (a) $a=3$ and $\Delta \mu_F^* \approx 1.25$. (b) $\Delta \mu_F=2 > \ln 4$ (dotted lines), $\Delta \mu_F=0.8<\ln 4$ (plain lines) and $a^* \approx 1.33$. A 3d plot is shown in \cite{SM}. } 
    \label{fig:phase_trans}
\end{figure}

{\it Phase transition -} When increasing $\Delta \mu_F$, we observe a smooth transition of the average error fraction $\overline{x}$ from random copies with error fraction $x_r$ (regions A and B) to accurate copies with error fraction $x_a$ (regions C and D). 
Since the difference between $\langle N_{x_a} \rangle $ and $\langle N_{x_r} \rangle$ grows exponentially with $L$ in regions B and C, this transition becomes sharp at $\Delta \mu_F = \Delta \mu_F^{*}$ when $L \to \infty$, as illustrated in \cref{fig:phase_trans}a. 
Therefore, for large $L$, accurate copies dominate and the average error fraction becomes $x_a$ when $\Delta \mu_F > \Delta \mu_F^{*}$ where $\Delta \mu_F^{*}$ depends on specificity $a$ and number of monomer types $m$. 
For any value of the specificity, accurate copies dominate in the large $L$ limit when
\begin{equation}
\label{eq:Landauer}
    \Delta \mu_F \geq \ln m \,.
\end{equation}
This condition can be interpreted as a Landauer's principle \cite{landauer_irreversibility_1961,andrieux_information_2013} for polymer copying: to copy information accurately, the per-monomer externally-provided free energy must be larger than the per-monomer entropy of configuration of the polymer $\ln (m^L) /L$ \cite{ouldridge_fundamental_2017}. We refer to $\Delta \mu_F=\ln m$ as the Landauer limit. 

Below the Landauer limit ($\Delta \mu_F < \ln m$) the average error fraction $\overline{x}$ decreases as specificity $a$ is increased at low specificity. If specificity $a$ crosses from below a threshold $a^*$ which depends on $\Delta \mu_F$ according to \cref{eq_muf*}, $\overline{x}$ jumps to the value $x_r$. For finite $L$, the transition becomes smooth (see \cref{fig:phase_trans}b). This transition occurs because increasing the specificity $a$ reduces the error fraction $x_a$ associated with accurate copies, but also slows down 
the templated assembly of these copies. 
Because the kinetics of the spontaneous process is independent of $a$, random copies will eventually dominate.
Above the Landauer limit ($\Delta \mu_F \geq \ln m$) the average error fraction $\overline{x}$ decreases as specificity $a$ is increased, but remains accurate and does not undergo a transition towards random copies.

In the large $L$ limit, the transition from random to accurate copies is a first order phase transition. The grey line shown in \cref{fig:phase_diag}a becomes a first order phase transition line in the limit of large $L$. 

{\it Population size -} 
We next ask if the number of copies that participate in this phase transition can vanish in the large $L$ limit. 
The number of accurate and random copies, $\langle N_{x_a} \rangle$ and $\langle N_{x_r} \rangle$, depend on the competition between the energetic and entropic contributions:
$\ln \langle N_{x} \rangle/L = \ln \lambda_{x L}/L + \ln \Omega_{x L}/L$.
If $\Delta \mu_r > \ln m$, the population of random copies $\langle N_{x_r} \rangle$ goes extinct for large $L$, with the energy difference $\ln \lambda_{x_r L}/L=-\Delta \mu_r$ and the entropy of configuration $\ln \Omega_{x_r L}/L= \ln m$. Similarly, if $\Delta \mu_F < \Delta \mu_r - \ln \lp 1+(m-1) e^{-a} \rp$, the population of accurate copies $\langle N_{x_a} \rangle$ vanishes, with $\ln \lambda_{x_a L}/L=-(\Delta \mu_r -\Delta \mu_F)-a x_a$ and $\ln \Omega_{x_a L}/L=\ln \lp 1+(m-1) e^{-a}\rp +a x_a$ \cite{SM}. 

We can now draw a phase diagram in the large $L$ limit as a function of $\Delta \mu_F$ and $\Delta \mu_r$, for a given specificity $a$, see \cref{fig:phase_diag}b. This diagram contains three regions: a region where accurate copies dominate, a region where random copies dominate, and a region where the population vanishes. The two boundary lines of the region of vanishing population are given by the two conditions of extinction discussed above. The boundary line between random and accurate copies occurs at $\Delta \mu_F = \Delta \mu_F^*$ where $\Delta \mu_F^*=\ln \lp m/(1+(m-1)e^{-a}) \rp$ in the large $L$ limit. The three regions meet at a triple point $(\Delta \mu_r = \ln m,\Delta \mu_F = \Delta \mu_F^*)$.

{\it Non equilibrium current -} We now investigate the non-equilibrium nature of the phase diagrams discussed above. In steady state, total assembly and disassembly are balanced for any sequence S: $k_a = k_d \langle N_S \rangle$. However the templated and spontaneous processes are not balanced individually, which is associated with a non-zero net average fuel current from the fuel bath to the waste bath $\langle J \rangle = L \sum_{j=1}^{m^L} (k_j^+ - \langle N_{S_j} \rangle k_j^-)$.
In the large $L$ limit $\langle J \rangle  \sim L k_0 \exp \lbk -\lp \Delta \mu_r - \Delta \mu_F - \ln \lp 1+ (m-1) e^{-a} \rp \rp L \rbk$ \cite{SM}, where $\sim$ describes asymptotic equality in the large $L$ limit. 

We now distinguish three regions of the phase diagram which differ in the transduction of fuel energy into useful information. When $\Delta \mu_F < \Delta \mu_r - \ln \lp 1+ (m-1) e^{-a} \rp$, the fuel current vanishes in the large $L$ limit. This region is delimited by the tilted line (both dashed and solid) in \cref{fig:phase_diag}b. In this case, no fuel is consumed and no accurate copies are produced.
The other two regions are located above this titled line, 
where the non-vanishing fuel current maintains the system in a non-equilibrium steady-state, 
and are delimited by the horizontal line in \cref{fig:phase_diag}b. 
If $\Delta \mu_F > \ln \lp m/ (1+(m-1)e^{-a}) \rp$, accurate copies dominate so fuel energy is efficiently converted into information, with $\langle J \rangle \sim L k_r \langle N_{x_a} \rangle$.
If $\Delta \mu_F < \ln \lp m/ (1+(m-1)e^{-a}) \rp$, random copies dominate in the large $L$ limit. In this case, fuel is burnt but no useful information is transmitted.

{\it Kinetic proofreading -} Fuel-driven error-correction mechanisms could increase copying accuracy by modifying the kinetics.
For example, kinetic proofreading feeds on fuel energy to undo copy errors at the expense of slowing the copy process \cite{hopfield_kinetic_1974,bennett_dissipation-error_1979,depken_intermittent_2013}. 
In our description of template copying, assembly and disassembly kinetics depend on sequence, but energy differences $\Delta \mu_r$ and $\Delta \mu_F$ do not.
Hence fuel-driven error-correction mechanisms could be implicitly accounted for in our model. 

Proofreading can take different pathways that can be associated with different amounts of fuel consumption. Hence, when explicitly coarse-graining over these pathways the micro-reversibility condition is broken \cite{ouldridge_thermodynamics_2017}
but effective kinetic rates can still be defined \cite{SM}. 
In general these rates are more complex than those introduced in the text after \cref{rct:copy_spont}. 
In some scenarios however, for example error correction by single state backtracking \cite{depken_intermittent_2013}, an effective kinetic prefactor, an effective specificity and an effective fuel free energy difference can be identified \cite{SM}. 
In such cases the error fraction $x_a$ of accurate copies will decrease, which in turn shifts the boundaries $a x_r$, $\Delta \mu_F^*$ and $a x_a$ of the phase diagram (\cref{fig:phase_diag}a), since these depend on $a$.

{\it Generalized reaction rates -} We now relax the assumption that the backward rates do not depend on the energies by introducing a fuel-dependent energy barrier in the templated rates: $k_j^+ =k_j e^{-(\Delta \mu_r - (1+\gamma) \Delta \mu_F)L} $ and $k_j^- =k_j e^{ \gamma \Delta \mu_F L} $ with $\gamma \in \mathbb{R}$. This changes the ratio of the time-scales associated with the templated and the spontaneous process in an $L$-dependent manner. 

\begin{figure}
	\centering
	\includegraphics[width=\linewidth]{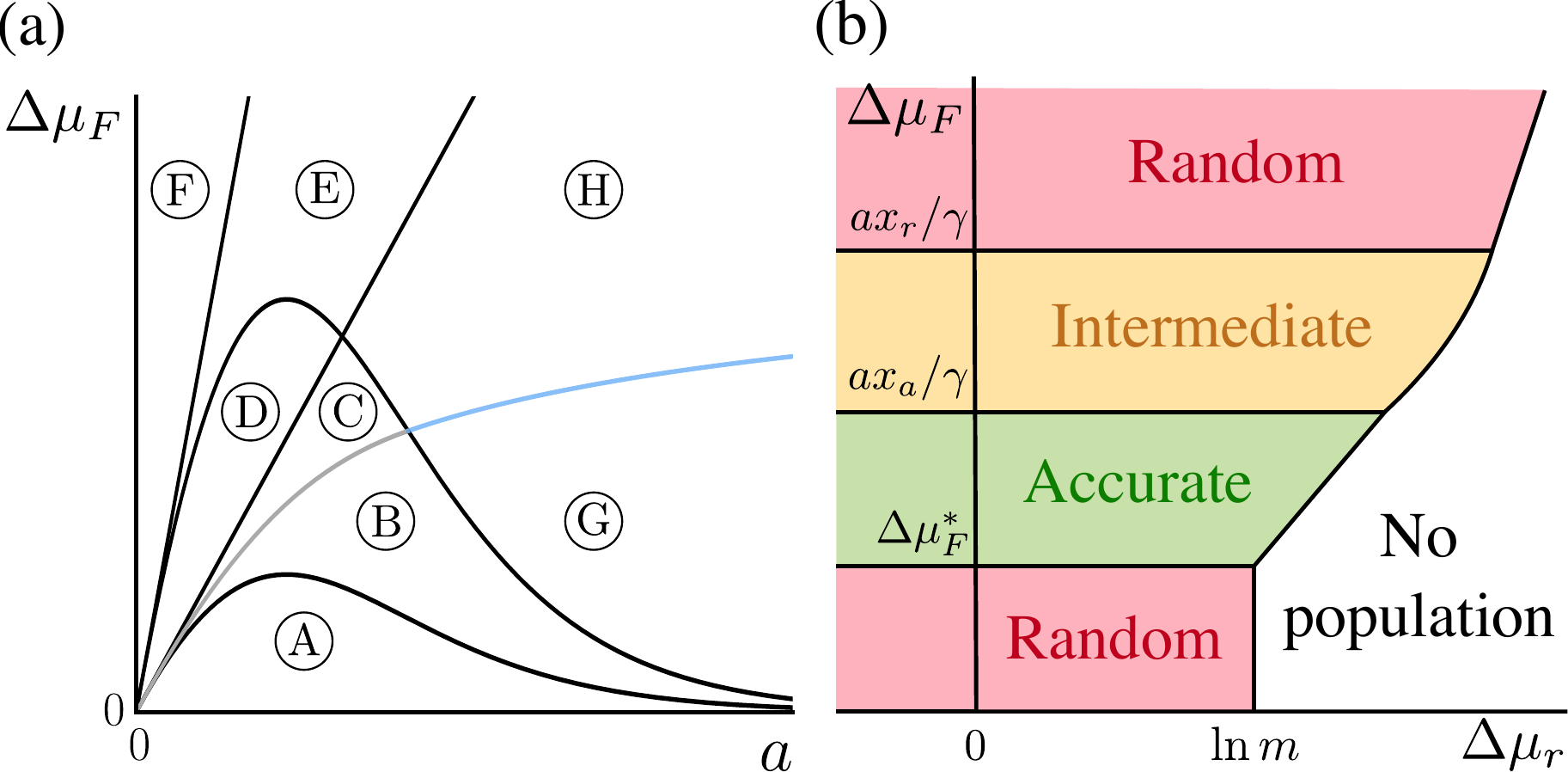}
	\caption{Schematic phase diagrams for the generalized reaction rates 
		and $\gamma >0$, for (a) finite $L$ and (b) large $L$ with a fixed value of $a$ such that $\ln \lp m/(1+(m-1)e^{-a}) \rp /(1+\gamma) < a x_a/\gamma$.}
	\label{fig:phase_diag_gam}
\end{figure}

We show in \cref{fig:phase_diag_gam} the finite and large $L$ phase diagrams for $\gamma>0$, which speeds up the kinetics of the templated process compared to $\gamma=0$, and allows templated disassembly to dominate over spontaneous disassembly for a range of error fractions $x$. 
The regions A-H and the extinction conditions are discussed in \cite{SM}. 
Similar to the example shown in \cref{fig:phase_diag}b, for small $\Delta \mu_F$, the system generates random copies. The transition from random to accurate copies occurs at the threshold $\Delta \mu_F^*$, which is reduced by a factor $1+\gamma$ relative to the value given in \cref{eq_muf*}.
Further increasing $\Delta \mu_F$ leads to a continuous increase in the average error fraction $\overline{x}=\gamma\Delta \mu_F/a$ (intermediate region) until the system re-enters a regime of random copying. 
This is because both assembly and disassembly are dominated by the templated process in this regime, which results in an equilibrium distribution of random copies.
There is thus a range of values $\Delta \mu_F^{*} < \Delta \mu_F < a x_a / \gamma$ where a maximum accuracy at error fraction $x_a$ is achieved.

{\it Discussion -} In this Letter we study populations of copolymer copies and their accuracy in a thermodynamically consistent template copying ensemble. We find sharp transitions between populations of random and accurate copies as a function of fuel driving and copying specificity.
Our coarse-grained approach reveals generic features of stochastic copying processes that are independent of many molecular details. It allows for an evaluation of the role of fuel driving, proofreading, and energy barriers on the population of accurate copies. We identify for given specificity the minimal cost of Gibbs free energy $\Delta \mu_F^*$ required to maintain a population of accurate copies. The minimal Gibbs free energy cost to be in a regime of accurate copying regardless of the specificity is given by the per monomer configurational entropy. In analogy to the Landauer principle of information erasure we refer to this as the Landauer limit. 

It will be interesting to extend this framework in several directions. First,
by allowing copies to serve as template themselves in an autocatalytic manner, 
perhaps offering new means to investigate the link between the statistical mechanics of replicating systems and the evolutionary process \cite{goldenfeld_life_2011,demetrius_boltzmann_2013,england_statistical_2013,genthon_branching_2022}. This might also allow for a re-investigation of Eigen's paradox of achieving high fidelity copies of large genomes \cite{eigen_selforganization_1971,leuthausser_statistical_1987}. More generally, other problems associated with the origin-of-life question could perhaps be freshly investigated within the template copying ensemble \cite{tkachenko_spontaneous_2015,matsubara_optimal_2016,matsubara_kinetic_2018,tkachenko_onset_2018,rosenberger_self-assembly_2021,kudella_structured_2021,goppel_thermodynamic_2022}.

Second, the free energy differences $L\Delta \mu_r$ and $L\Delta \mu_F$ might not scale with template length $L$ and the specificity $a$ may depend on the error fraction $x$, as observed for example in kinetic proofreading \cite{SM}. Hence, the kinetic rates defined in this work are the first order terms in the Taylor expansions of more general rates. 
In principle, next order terms could be systematically determined in future work.

Third, in the one-step copy process discussed above 
intermediate stages of polymer growth 
are not represented. It is an interesting direction of future work to refine our description and take into account such intermediate states.

It will be interesting to test our results in experiments. For example template copying could be realized with in vitro transcription assays where a DNA template is used to generate RNA copies \cite{losick_vitro_1972}. Furthermore,  
Polymerase-Chain-Reaction (PCR) experiments using primers that attach to only one of the two strands of the DNA template
would generate complementary DNA copies of the template strand and provide another example of template copying \cite{mullis_specific_1986}. Performing such DNA copying experiments and measuring distributions of error fractions in the produced sequences could provide experimental tests of our results and infer in which regimes these processes operate. 

While inspired by the copy process of biological DNA polymers \cite{SM}, our formalism more generally applies to systems of information transfer from a template to a copy of the template. For example, our approach equally applies to the copy process from RNA to protein (translation, with 20 ‘monomer' types), or more generally for copy processes with any number of monomer types. 
Our formalism allows for the discussion of a trade-off that arises when increasing the number $m$ of monomer types, for large $L$. The energetic cost required to be in a regime of accurate copying is high for low values of $m$ as compared to large values of $m$, but the accuracy of copies is higher. This is because an increase of $m$ allows for the use of shorter polymers for encoding the same amount of information $\Omega=m^L$ , while the more complex encoding is more prone to mistakes (\cref{eq_xa}). Indeed, the minimal energy to copy a sequence with information content $\Omega$ with specificity $a$ is given by $E_{\rm{tot}}^* = L(m) \Delta \mu_F^* \sim \lp 1- \ln \lp 1+(m-1)e^{-a} \rp/\ln m \rp \ln \Omega$, which decreases with $m$. It is interesting to speculate that this cost-accuracy trade-off is relevant from an evolutionary point of view, manifested in the choice of $m=4$ for maintaining the genome in DNA form at high fidelity, and for copying genomic information to protein peptide sequences with $m=20$ at lower energetic costs and reduced requirements on fidelity.

{\it Acknowledgments -} We thank B. Qureshi and T. E. Ouldridge for a critical reading of our manuscript and for providing a preprint of their work \cite{qureshi_information_2024}.


\begin{thebibliography}{43}%
	\makeatletter
	\providecommand \@ifxundefined [1]{%
		\@ifx{#1\undefined}
	}%
	\providecommand \@ifnum [1]{%
		\ifnum #1\expandafter \@firstoftwo
		\else \expandafter \@secondoftwo
		\fi
	}%
	\providecommand \@ifx [1]{%
		\ifx #1\expandafter \@firstoftwo
		\else \expandafter \@secondoftwo
		\fi
	}%
	\providecommand \natexlab [1]{#1}%
	\providecommand \enquote  [1]{``#1''}%
	\providecommand \bibnamefont  [1]{#1}%
	\providecommand \bibfnamefont [1]{#1}%
	\providecommand \citenamefont [1]{#1}%
	\providecommand \href@noop [0]{\@secondoftwo}%
	\providecommand \href [0]{\begingroup \@sanitize@url \@href}%
	\providecommand \@href[1]{\@@startlink{#1}\@@href}%
	\providecommand \@@href[1]{\endgroup#1\@@endlink}%
	\providecommand \@sanitize@url [0]{\catcode `\\12\catcode `\$12\catcode
		`\&12\catcode `\#12\catcode `\^12\catcode `\_12\catcode `\%12\relax}%
	\providecommand \@@startlink[1]{}%
	\providecommand \@@endlink[0]{}%
	\providecommand \url  [0]{\begingroup\@sanitize@url \@url }%
	\providecommand \@url [1]{\endgroup\@href {#1}{\urlprefix }}%
	\providecommand \urlprefix  [0]{URL }%
	\providecommand \Eprint [0]{\href }%
	\providecommand \doibase [0]{https://doi.org/}%
	\providecommand \selectlanguage [0]{\@gobble}%
	\providecommand \bibinfo  [0]{\@secondoftwo}%
	\providecommand \bibfield  [0]{\@secondoftwo}%
	\providecommand \translation [1]{[#1]}%
	\providecommand \BibitemOpen [0]{}%
	\providecommand \bibitemStop [0]{}%
	\providecommand \bibitemNoStop [0]{.\EOS\space}%
	\providecommand \EOS [0]{\spacefactor3000\relax}%
	\providecommand \BibitemShut  [1]{\csname bibitem#1\endcsname}%
	\let\auto@bib@innerbib\@empty
	\bibitem [{\citenamefont {Kornberg}\ and\ \citenamefont
		{Baker}(1992)}]{kornberg_dna_1992}%
	\BibitemOpen
	\bibfield  {author} {\bibinfo {author} {\bibfnamefont {A.}~\bibnamefont
			{Kornberg}}\ and\ \bibinfo {author} {\bibfnamefont {T.~A.}\ \bibnamefont
			{Baker}},\ }\href@noop {} {\emph {\bibinfo {title} {{DNA} {Replication}}}},\
	\bibinfo {edition} {2nd}\ ed.\ (\bibinfo  {publisher} {W. H. Freeman and
		Company},\ \bibinfo {address} {New York},\ \bibinfo {year}
	{1992})\BibitemShut {NoStop}%
	\bibitem [{\citenamefont {Lieberman}\ \emph {et~al.}(1955)\citenamefont
		{Lieberman}, \citenamefont {Kornberg},\ and\ \citenamefont
		{Simms}}]{lieberman_enzymatic_1955}%
	\BibitemOpen
	\bibfield  {author} {\bibinfo {author} {\bibfnamefont {I.}~\bibnamefont
			{Lieberman}}, \bibinfo {author} {\bibfnamefont {A.}~\bibnamefont
			{Kornberg}},\ and\ \bibinfo {author} {\bibfnamefont {E.~S.}\ \bibnamefont
			{Simms}},\ }\bibfield  {title} {{\selectlanguage {en}\bibinfo {title}
			{Enzymatic synthesis of nucleoside diphosphates and triphosphates}},\ }\href
	{https://doi.org/10.1016/S0021-9258(18)66050-8} {\bibfield  {journal}
		{\bibinfo  {journal} {J. Biol. Chem.}\ }\textbf {\bibinfo {volume} {215}},\
		\bibinfo {pages} {429} (\bibinfo {year} {1955})}\BibitemShut {NoStop}%
	\bibitem [{\citenamefont {Yarranton}\ and\ \citenamefont
		{Gefter}(1979)}]{yarranton_enzyme-catalyzed_1979}%
	\BibitemOpen
	\bibfield  {author} {\bibinfo {author} {\bibfnamefont {G.~T.}\ \bibnamefont
			{Yarranton}}\ and\ \bibinfo {author} {\bibfnamefont {M.~L.}\ \bibnamefont
			{Gefter}},\ }\bibfield  {title} {{\selectlanguage {en}\bibinfo {title}
			{Enzyme-catalyzed {DNA} unwinding: {Studies} on \textit{{Escherichia} coli
					rep} protein}},\ }\href {https://doi.org/10.1073/pnas.76.4.1658} {\bibfield
		{journal} {\bibinfo  {journal} {Proc. Natl. Acad. Sci. U.S.A.}\ }\textbf
		{\bibinfo {volume} {76}},\ \bibinfo {pages} {1658} (\bibinfo {year}
		{1979})}\BibitemShut {NoStop}%
	\bibitem [{\citenamefont {Lehman}\ \emph {et~al.}(1958)\citenamefont {Lehman},
		\citenamefont {Bessman}, \citenamefont {Simms},\ and\ \citenamefont
		{Kornberg}}]{lehman_enzymatic_1958}%
	\BibitemOpen
	\bibfield  {author} {\bibinfo {author} {\bibfnamefont {I.}~\bibnamefont
			{Lehman}}, \bibinfo {author} {\bibfnamefont {M.~J.}\ \bibnamefont {Bessman}},
		\bibinfo {author} {\bibfnamefont {E.~S.}\ \bibnamefont {Simms}},\ and\
		\bibinfo {author} {\bibfnamefont {A.}~\bibnamefont {Kornberg}},\ }\bibfield
	{title} {{\selectlanguage {en}\bibinfo {title} {Enzymatic {Synthesis} of
				{Deoxyribonucleic} {Acid}}},\ }\href
	{https://doi.org/10.1016/S0021-9258(19)68048-8} {\bibfield  {journal}
		{\bibinfo  {journal} {J. Biol. Chem.}\ }\textbf {\bibinfo {volume} {233}},\
		\bibinfo {pages} {163} (\bibinfo {year} {1958})}\BibitemShut {NoStop}%
	\bibitem [{\citenamefont {Yang}(2011)}]{yang_nucleases_2011}%
	\BibitemOpen
	\bibfield  {author} {\bibinfo {author} {\bibfnamefont {W.}~\bibnamefont
			{Yang}},\ }\bibfield  {title} {{\selectlanguage {en}\bibinfo {title}
			{Nucleases: diversity of structure, function and mechanism}},\ }\href
	{https://doi.org/10.1017/S0033583510000181} {\bibfield  {journal} {\bibinfo
			{journal} {Quart. Rev. Biophys.}\ }\textbf {\bibinfo {volume} {44}},\
		\bibinfo {pages} {1} (\bibinfo {year} {2011})}\BibitemShut {NoStop}%
	\bibitem [{\citenamefont {Andrieux}\ and\ \citenamefont
		{Gaspard}(2008)}]{andrieux_nonequilibrium_2008}%
	\BibitemOpen
	\bibfield  {author} {\bibinfo {author} {\bibfnamefont {D.}~\bibnamefont
			{Andrieux}}\ and\ \bibinfo {author} {\bibfnamefont {P.}~\bibnamefont
			{Gaspard}},\ }\bibfield  {title} {{\selectlanguage {en}\bibinfo {title}
			{Nonequilibrium generation of information in copolymerization processes}},\
	}\href {https://doi.org/10.1073/pnas.0802049105} {\bibfield  {journal}
		{\bibinfo  {journal} {Proc. Natl. Acad. Sci. U.S.A.}\ }\textbf {\bibinfo
			{volume} {105}},\ \bibinfo {pages} {9516} (\bibinfo {year}
		{2008})}\BibitemShut {NoStop}%
	\bibitem [{\citenamefont {Andrieux}\ and\ \citenamefont
		{Gaspard}(2009)}]{andrieux_molecular_2009}%
	\BibitemOpen
	\bibfield  {author} {\bibinfo {author} {\bibfnamefont {D.}~\bibnamefont
			{Andrieux}}\ and\ \bibinfo {author} {\bibfnamefont {P.}~\bibnamefont
			{Gaspard}},\ }\bibfield  {title} {{\selectlanguage {en}\bibinfo {title}
			{Molecular information processing in nonequilibrium copolymerizations}},\
	}\href {https://doi.org/10.1063/1.3050099} {\bibfield  {journal} {\bibinfo
			{journal} {J. Chem. Phys.}\ }\textbf {\bibinfo {volume} {130}},\ \bibinfo
		{pages} {014901} (\bibinfo {year} {2009})}\BibitemShut {NoStop}%
	\bibitem [{\citenamefont {Sartori}\ and\ \citenamefont
		{Pigolotti}(2013)}]{sartori_kinetic_2013}%
	\BibitemOpen
	\bibfield  {author} {\bibinfo {author} {\bibfnamefont {P.}~\bibnamefont
			{Sartori}}\ and\ \bibinfo {author} {\bibfnamefont {S.}~\bibnamefont
			{Pigolotti}},\ }\bibfield  {title} {{\selectlanguage {en}\bibinfo {title}
			{Kinetic versus {Energetic} {Discrimination} in {Biological} {Copying}}},\
	}\href {https://doi.org/10.1103/PhysRevLett.110.188101} {\bibfield  {journal}
		{\bibinfo  {journal} {Phys. Rev. Lett.}\ }\textbf {\bibinfo {volume} {110}},\
		\bibinfo {pages} {188101} (\bibinfo {year} {2013})}\BibitemShut {NoStop}%
	\bibitem [{\citenamefont {Gaspard}\ and\ \citenamefont
		{Andrieux}(2014)}]{gaspard_kinetics_2014}%
	\BibitemOpen
	\bibfield  {author} {\bibinfo {author} {\bibfnamefont {P.}~\bibnamefont
			{Gaspard}}\ and\ \bibinfo {author} {\bibfnamefont {D.}~\bibnamefont
			{Andrieux}},\ }\bibfield  {title} {{\selectlanguage {en}\bibinfo {title}
			{Kinetics and thermodynamics of first-order {Markov} chain
				copolymerization}},\ }\href {https://doi.org/10.1063/1.4890821} {\bibfield
		{journal} {\bibinfo  {journal} {J. Chem. Phys.}\ }\textbf {\bibinfo {volume}
			{141}},\ \bibinfo {pages} {044908} (\bibinfo {year} {2014})}\BibitemShut
	{NoStop}%
	\bibitem [{\citenamefont {Sartori}\ and\ \citenamefont
		{Pigolotti}(2015)}]{sartori_thermodynamics_2015}%
	\BibitemOpen
	\bibfield  {author} {\bibinfo {author} {\bibfnamefont {P.}~\bibnamefont
			{Sartori}}\ and\ \bibinfo {author} {\bibfnamefont {S.}~\bibnamefont
			{Pigolotti}},\ }\bibfield  {title} {{\selectlanguage {en}\bibinfo {title}
			{Thermodynamics of {Error} {Correction}}},\ }\href
	{https://doi.org/10.1103/PhysRevX.5.041039} {\bibfield  {journal} {\bibinfo
			{journal} {Phys. Rev. X}\ }\textbf {\bibinfo {volume} {5}},\ \bibinfo {pages}
		{041039} (\bibinfo {year} {2015})}\BibitemShut {NoStop}%
	\bibitem [{\citenamefont {Rao}\ and\ \citenamefont
		{Peliti}(2015)}]{rao_thermodynamics_2015}%
	\BibitemOpen
	\bibfield  {author} {\bibinfo {author} {\bibfnamefont {R.}~\bibnamefont
			{Rao}}\ and\ \bibinfo {author} {\bibfnamefont {L.}~\bibnamefont {Peliti}},\
	}\bibfield  {title} {{\selectlanguage {en}\bibinfo {title} {Thermodynamics of
				accuracy in kinetic proofreading: dissipation and efficiency trade-offs}},\
	}\href {https://doi.org/10.1088/1742-5468/2015/06/P06001} {\bibfield
		{journal} {\bibinfo  {journal} {J. Stat. Mech.}\ }\textbf {\bibinfo {volume}
			{2015}},\ \bibinfo {pages} {P06001} (\bibinfo {year} {2015})}\BibitemShut
	{NoStop}%
	\bibitem [{\citenamefont {Ouldridge}\ and\ \citenamefont {Rein~ten
			Wolde}(2017)}]{ouldridge_fundamental_2017}%
	\BibitemOpen
	\bibfield  {author} {\bibinfo {author} {\bibfnamefont {T.~E.}\ \bibnamefont
			{Ouldridge}}\ and\ \bibinfo {author} {\bibfnamefont {P.}~\bibnamefont
			{Rein~ten Wolde}},\ }\bibfield  {title} {{\selectlanguage {en}\bibinfo
			{title} {Fundamental {Costs} in the {Production} and {Destruction} of
				{Persistent} {Polymer} {Copies}}},\ }\href
	{https://doi.org/10.1103/PhysRevLett.118.158103} {\bibfield  {journal}
		{\bibinfo  {journal} {Phys. Rev. Lett.}\ }\textbf {\bibinfo {volume} {118}},\
		\bibinfo {pages} {158103} (\bibinfo {year} {2017})}\BibitemShut {NoStop}%
	\bibitem [{\citenamefont {Poulton}\ \emph {et~al.}(2019)\citenamefont
		{Poulton}, \citenamefont {Rein~ten Wolde},\ and\ \citenamefont
		{Ouldridge}}]{poulton_nonequilibrium_2019}%
	\BibitemOpen
	\bibfield  {author} {\bibinfo {author} {\bibfnamefont {J.~M.}\ \bibnamefont
			{Poulton}}, \bibinfo {author} {\bibfnamefont {P.}~\bibnamefont {Rein~ten
				Wolde}},\ and\ \bibinfo {author} {\bibfnamefont {T.~E.}\ \bibnamefont
			{Ouldridge}},\ }\bibfield  {title} {{\selectlanguage {en}\bibinfo {title}
			{Nonequilibrium correlations in minimal dynamical models of polymer
				copying}},\ }\href {https://doi.org/10.1073/pnas.1808775116} {\bibfield
		{journal} {\bibinfo  {journal} {Proc. Natl. Acad. Sci. U.S.A.}\ }\textbf
		{\bibinfo {volume} {116}},\ \bibinfo {pages} {1946} (\bibinfo {year}
		{2019})}\BibitemShut {NoStop}%
	\bibitem [{\citenamefont {Chiuchiú}\ \emph {et~al.}(2019)\citenamefont
		{Chiuchiú}, \citenamefont {Tu},\ and\ \citenamefont
		{Pigolotti}}]{chiuchiu_error-speed_2019}%
	\BibitemOpen
	\bibfield  {author} {\bibinfo {author} {\bibfnamefont {D.}~\bibnamefont
			{Chiuchiú}}, \bibinfo {author} {\bibfnamefont {Y.}~\bibnamefont {Tu}},\ and\
		\bibinfo {author} {\bibfnamefont {S.}~\bibnamefont {Pigolotti}},\ }\bibfield
	{title} {{\selectlanguage {en}\bibinfo {title} {Error-{Speed} {Correlations}
				in {Biopolymer} {Synthesis}}},\ }\href
	{https://doi.org/10.1103/PhysRevLett.123.038101} {\bibfield  {journal}
		{\bibinfo  {journal} {Phys. Rev. Lett.}\ }\textbf {\bibinfo {volume} {123}},\
		\bibinfo {pages} {038101} (\bibinfo {year} {2019})}\BibitemShut {NoStop}%
	\bibitem [{\citenamefont {Sahoo}\ \emph {et~al.}(2021)\citenamefont {Sahoo},
		\citenamefont {Noushad}, \citenamefont {Baral},\ and\ \citenamefont
		{Klumpp}}]{sahoo_accuracy_2021}%
	\BibitemOpen
	\bibfield  {author} {\bibinfo {author} {\bibfnamefont {M.}~\bibnamefont
			{Sahoo}}, \bibinfo {author} {\bibfnamefont {A.}~\bibnamefont {Noushad}},
		\bibinfo {author} {\bibfnamefont {P.~R.}\ \bibnamefont {Baral}},\ and\
		\bibinfo {author} {\bibfnamefont {S.}~\bibnamefont {Klumpp}},\ }\bibfield
	{title} {{\selectlanguage {en}\bibinfo {title} {Accuracy and speed of
				elongation in a minimal model of {DNA} replication}},\ }\href
	{https://doi.org/10.1103/PhysRevE.104.034417} {\bibfield  {journal} {\bibinfo
			{journal} {Phys. Rev. E}\ }\textbf {\bibinfo {volume} {104}},\ \bibinfo
		{pages} {034417} (\bibinfo {year} {2021})}\BibitemShut {NoStop}%
	\bibitem [{\citenamefont {Poulton}\ and\ \citenamefont
		{Ouldridge}(2021)}]{poulton_edge-effects_2021}%
	\BibitemOpen
	\bibfield  {author} {\bibinfo {author} {\bibfnamefont {J.~M.}\ \bibnamefont
			{Poulton}}\ and\ \bibinfo {author} {\bibfnamefont {T.~E.}\ \bibnamefont
			{Ouldridge}},\ }\bibfield  {title} {{\selectlanguage {en}\bibinfo {title}
			{Edge-effects dominate copying thermodynamics for finite-length molecular
				oligomers}},\ }\href {https://doi.org/10.1088/1367-2630/ac0389} {\bibfield
		{journal} {\bibinfo  {journal} {New J. Phys.}\ }\textbf {\bibinfo {volume}
			{23}},\ \bibinfo {pages} {063061} (\bibinfo {year} {2021})}\BibitemShut
	{NoStop}%
	\bibitem [{\citenamefont {Juritz}\ \emph {et~al.}(2022)\citenamefont {Juritz},
		\citenamefont {Poulton},\ and\ \citenamefont
		{Ouldridge}}]{juritz_minimal_2022}%
	\BibitemOpen
	\bibfield  {author} {\bibinfo {author} {\bibfnamefont {J.}~\bibnamefont
			{Juritz}}, \bibinfo {author} {\bibfnamefont {J.~M.}\ \bibnamefont
			{Poulton}},\ and\ \bibinfo {author} {\bibfnamefont {T.~E.}\ \bibnamefont
			{Ouldridge}},\ }\bibfield  {title} {{\selectlanguage {en}\bibinfo {title}
			{Minimal mechanism for cyclic templating of length-controlled copolymers
				under isothermal conditions}},\ }\href {https://doi.org/10.1063/5.0077865}
	{\bibfield  {journal} {\bibinfo  {journal} {J. Chem. Phys.}\ }\textbf
		{\bibinfo {volume} {156}},\ \bibinfo {pages} {074103} (\bibinfo {year}
		{2022})}\BibitemShut {NoStop}%
	\bibitem [{\citenamefont {Bennett}(1979)}]{bennett_dissipation-error_1979}%
	\BibitemOpen
	\bibfield  {author} {\bibinfo {author} {\bibfnamefont {C.~H.}\ \bibnamefont
			{Bennett}},\ }\bibfield  {title} {{\selectlanguage {en}\bibinfo {title}
			{Dissipation-error tradeoff in proofreading}},\ }\href
	{https://doi.org/10.1016/0303-2647(79)90003-0} {\bibfield  {journal}
		{\bibinfo  {journal} {Biosystems}\ }\textbf {\bibinfo {volume} {11}},\
		\bibinfo {pages} {85} (\bibinfo {year} {1979})}\BibitemShut {NoStop}%
	\bibitem [{qur()}]{qureshi_information_2024}%
	\BibitemOpen
	\href@noop {} {}\bibinfo {note} {We would like to point the reader to the
		preprint B. Qureshi, J. M. Poulton and T. E. Ouldridge, arXiv:2404.02791v1
		that became available during the review stage of this manuscript, which
		discusses thermodynamic limits on copying accuracy.}\BibitemShut {Stop}%
	\bibitem [{SM()}]{SM}%
	\BibitemOpen
	\href@noop {} {}\bibinfo {note} {See Supplemental Material at [URL will be
		inserted by publisher] for additional derivations, which includes Refs. [21-23].}\BibitemShut {Stop}%
	\bibitem [{\citenamefont {Goodman}\ \emph {et~al.}(1993)\citenamefont
		{Goodman}, \citenamefont {Creighton}, \citenamefont {Bloom}, \citenamefont
		{Petruska},\ and\ \citenamefont {Kunkel}}]{goodman_biochemical_1993}%
	\BibitemOpen
	\bibfield  {author} {\bibinfo {author} {\bibfnamefont {M.~F.}\ \bibnamefont
			{Goodman}}, \bibinfo {author} {\bibfnamefont {S.}~\bibnamefont {Creighton}},
		\bibinfo {author} {\bibfnamefont {L.~B.}\ \bibnamefont {Bloom}}, \bibinfo
		{author} {\bibfnamefont {J.}~\bibnamefont {Petruska}},\ and\ \bibinfo
		{author} {\bibfnamefont {T.~A.}\ \bibnamefont {Kunkel}},\ }\bibfield  {title}
	{{\selectlanguage {en}\bibinfo {title} {Biochemical {Basis} of {DNA}
				{Replication} {Fidelity}}},\ }\href
	{https://doi.org/10.3109/10409239309086792} {\bibfield  {journal} {\bibinfo
			{journal} {Crit. Rev. Biochem. Mol. Biol.}\ }\textbf {\bibinfo {volume}
			{28}},\ \bibinfo {pages} {83} (\bibinfo {year} {1993})}\BibitemShut {NoStop}%
	\bibitem [{\citenamefont {Thomas}\ \emph {et~al.}(1998)\citenamefont {Thomas},
		\citenamefont {Platas},\ and\ \citenamefont
		{Hawley}}]{thomas_transcriptional_1998}%
	\BibitemOpen
	\bibfield  {author} {\bibinfo {author} {\bibfnamefont {M.~J.}\ \bibnamefont
			{Thomas}}, \bibinfo {author} {\bibfnamefont {A.~A.}\ \bibnamefont {Platas}},\
		and\ \bibinfo {author} {\bibfnamefont {D.~K.}\ \bibnamefont {Hawley}},\
	}\bibfield  {title} {{\selectlanguage {en}\bibinfo {title} {Transcriptional
				{Fidelity} and {Proofreading} by {RNA} {Polymerase} {II}}},\ }\href
	{https://doi.org/10.1016/S0092-8674(00)81191-5} {\bibfield  {journal}
		{\bibinfo  {journal} {Cell}\ }\textbf {\bibinfo {volume} {93}},\ \bibinfo
		{pages} {627} (\bibinfo {year} {1998})}\BibitemShut {NoStop}%
	\bibitem [{\citenamefont {Gillespie}(1977)}]{gillespie_exact_1977}%
	\BibitemOpen
	\bibfield  {author} {\bibinfo {author} {\bibfnamefont {D.~T.}\ \bibnamefont
			{Gillespie}},\ }\bibfield  {title} {{\selectlanguage {en}\bibinfo {title}
			{Exact stochastic simulation of coupled chemical reactions}},\ }\href
	{https://doi.org/10.1021/j100540a008} {\bibfield  {journal} {\bibinfo
			{journal} {J. Phys. Chem.}\ }\textbf {\bibinfo {volume} {81}},\ \bibinfo
		{pages} {2340} (\bibinfo {year} {1977})}\BibitemShut {NoStop}%
	\bibitem [{\citenamefont {Landauer}(1961)}]{landauer_irreversibility_1961}%
	\BibitemOpen
	\bibfield  {author} {\bibinfo {author} {\bibfnamefont {R.}~\bibnamefont
			{Landauer}},\ }\bibfield  {title} {{\selectlanguage {en}\bibinfo {title}
			{Irreversibility and {Heat} {Generation} in the {Computing} {Process}}},\
	}\href {https://doi.org/10.1147/rd.53.0183} {\bibfield  {journal} {\bibinfo
			{journal} {IBM J. Res. Dev.}\ }\textbf {\bibinfo {volume} {5}},\ \bibinfo
		{pages} {183} (\bibinfo {year} {1961})}\BibitemShut {NoStop}%
	\bibitem [{\citenamefont {Andrieux}\ and\ \citenamefont
		{Gaspard}(2013)}]{andrieux_information_2013}%
	\BibitemOpen
	\bibfield  {author} {\bibinfo {author} {\bibfnamefont {D.}~\bibnamefont
			{Andrieux}}\ and\ \bibinfo {author} {\bibfnamefont {P.}~\bibnamefont
			{Gaspard}},\ }\bibfield  {title} {{\selectlanguage {en}\bibinfo {title}
			{Information erasure in copolymers}},\ }\href
	{https://doi.org/10.1209/0295-5075/103/30004} {\bibfield  {journal} {\bibinfo
			{journal} {EPL}\ }\textbf {\bibinfo {volume} {103}},\ \bibinfo {pages}
		{30004} (\bibinfo {year} {2013})}\BibitemShut {NoStop}%
	\bibitem [{\citenamefont {Hopfield}(1974)}]{hopfield_kinetic_1974}%
	\BibitemOpen
	\bibfield  {author} {\bibinfo {author} {\bibfnamefont {J.~J.}\ \bibnamefont
			{Hopfield}},\ }\bibfield  {title} {{\selectlanguage {en}\bibinfo {title}
			{Kinetic {Proofreading}: {A} {New} {Mechanism} for {Reducing} {Errors} in
				{Biosynthetic} {Processes} {Requiring} {High} {Specificity}}},\ }\href
	{https://doi.org/10.1073/pnas.71.10.4135} {\bibfield  {journal} {\bibinfo
			{journal} {Proc. Natl. Acad. Sci. U.S.A.}\ }\textbf {\bibinfo {volume}
			{71}},\ \bibinfo {pages} {4135} (\bibinfo {year} {1974})}\BibitemShut
	{NoStop}%
	\bibitem [{\citenamefont {Depken}\ \emph {et~al.}(2013)\citenamefont {Depken},
		\citenamefont {Parrondo},\ and\ \citenamefont
		{Grill}}]{depken_intermittent_2013}%
	\BibitemOpen
	\bibfield  {author} {\bibinfo {author} {\bibfnamefont {M.}~\bibnamefont
			{Depken}}, \bibinfo {author} {\bibfnamefont {J.~M.}\ \bibnamefont
			{Parrondo}},\ and\ \bibinfo {author} {\bibfnamefont {S.~W.}\ \bibnamefont
			{Grill}},\ }\bibfield  {title} {{\selectlanguage {en}\bibinfo {title}
			{Intermittent {Transcription} {Dynamics} for the {Rapid} {Production} of
				{Long} {Transcripts} of {High} {Fidelity}}},\ }\href
	{https://doi.org/10.1016/j.celrep.2013.09.007} {\bibfield  {journal}
		{\bibinfo  {journal} {Cell Rep.}\ }\textbf {\bibinfo {volume} {5}},\ \bibinfo
		{pages} {521} (\bibinfo {year} {2013})}\BibitemShut {NoStop}%
	\bibitem [{\citenamefont {Ouldridge}\ \emph {et~al.}(2017)\citenamefont
		{Ouldridge}, \citenamefont {Govern},\ and\ \citenamefont
		{Ten~Wolde}}]{ouldridge_thermodynamics_2017}%
	\BibitemOpen
	\bibfield  {author} {\bibinfo {author} {\bibfnamefont {T.~E.}\ \bibnamefont
			{Ouldridge}}, \bibinfo {author} {\bibfnamefont {C.~C.}\ \bibnamefont
			{Govern}},\ and\ \bibinfo {author} {\bibfnamefont {P.~R.}\ \bibnamefont
			{Ten~Wolde}},\ }\bibfield  {title} {{\selectlanguage {en}\bibinfo {title}
			{Thermodynamics of {Computational} {Copying} in {Biochemical} {Systems}}},\
	}\href {https://doi.org/10.1103/PhysRevX.7.021004} {\bibfield  {journal}
		{\bibinfo  {journal} {Phys. Rev. X}\ }\textbf {\bibinfo {volume} {7}},\
		\bibinfo {pages} {021004} (\bibinfo {year} {2017})}\BibitemShut {NoStop}%
	\bibitem [{\citenamefont {Goldenfeld}\ and\ \citenamefont
		{Woese}(2011)}]{goldenfeld_life_2011}%
	\BibitemOpen
	\bibfield  {author} {\bibinfo {author} {\bibfnamefont {N.}~\bibnamefont
			{Goldenfeld}}\ and\ \bibinfo {author} {\bibfnamefont {C.}~\bibnamefont
			{Woese}},\ }\bibfield  {title} {\bibinfo {title} {Life is physics: Evolution
			as a collective phenomenon far from equilibrium},\ }\href
	{https://doi.org/10.1146/annurev-conmatphys-062910-140509} {\bibfield
		{journal} {\bibinfo  {journal} {Annu. Rev. Condens. Matter Phys.}\ }\textbf
		{\bibinfo {volume} {2}},\ \bibinfo {pages} {375} (\bibinfo {year}
		{2011})}\BibitemShut {NoStop}%
	\bibitem [{\citenamefont {Demetrius}(2013)}]{demetrius_boltzmann_2013}%
	\BibitemOpen
	\bibfield  {author} {\bibinfo {author} {\bibfnamefont {L.~A.}\ \bibnamefont
			{Demetrius}},\ }\bibfield  {title} {{\selectlanguage {en}\bibinfo {title}
			{Boltzmann, {Darwin} and {Directionality} theory}},\ }\href
	{https://doi.org/10.1016/j.physrep.2013.04.001} {\bibfield  {journal}
		{\bibinfo  {journal} {Phys. Rep.}\ }\textbf {\bibinfo {volume} {530}},\
		\bibinfo {pages} {1} (\bibinfo {year} {2013})}\BibitemShut {NoStop}%
	\bibitem [{\citenamefont {England}(2013)}]{england_statistical_2013}%
	\BibitemOpen
	\bibfield  {author} {\bibinfo {author} {\bibfnamefont {J.~L.}\ \bibnamefont
			{England}},\ }\bibfield  {title} {{\selectlanguage {en}\bibinfo {title}
			{Statistical physics of self-replication}},\ }\href
	{https://doi.org/10.1063/1.4818538} {\bibfield  {journal} {\bibinfo
			{journal} {J. Chem. Phys.}\ }\textbf {\bibinfo {volume} {139}},\ \bibinfo
		{pages} {121923} (\bibinfo {year} {2013})}\BibitemShut {NoStop}%
	\bibitem [{\citenamefont {Genthon}\ \emph {et~al.}(2022)\citenamefont
		{Genthon}, \citenamefont {García-García},\ and\ \citenamefont
		{Lacoste}}]{genthon_branching_2022}%
	\BibitemOpen
	\bibfield  {author} {\bibinfo {author} {\bibfnamefont {A.}~\bibnamefont
			{Genthon}}, \bibinfo {author} {\bibfnamefont {R.}~\bibnamefont
			{García-García}},\ and\ \bibinfo {author} {\bibfnamefont {D.}~\bibnamefont
			{Lacoste}},\ }\bibfield  {title} {{\selectlanguage {en}\bibinfo {title}
			{Branching processes with resetting as a model for cell division}},\ }\href
	{https://doi.org/10.1088/1751-8121/ac491a} {\bibfield  {journal} {\bibinfo
			{journal} {J. Phys. A: Math. Theor.}\ }\textbf {\bibinfo {volume} {55}},\
		\bibinfo {pages} {074001} (\bibinfo {year} {2022})}\BibitemShut {NoStop}%
	\bibitem [{\citenamefont {Eigen}(1971)}]{eigen_selforganization_1971}%
	\BibitemOpen
	\bibfield  {author} {\bibinfo {author} {\bibfnamefont {M.}~\bibnamefont
			{Eigen}},\ }\bibfield  {title} {{\selectlanguage {en}\bibinfo {title}
			{Selforganization of matter and the evolution of biological
				macromolecules}},\ }\href {https://doi.org/10.1007/BF00623322} {\bibfield
		{journal} {\bibinfo  {journal} {Naturwissenschaften}\ }\textbf {\bibinfo
			{volume} {58}},\ \bibinfo {pages} {465} (\bibinfo {year} {1971})}\BibitemShut
	{NoStop}%
	\bibitem [{\citenamefont {Leuthäusser}(1987)}]{leuthausser_statistical_1987}%
	\BibitemOpen
	\bibfield  {author} {\bibinfo {author} {\bibfnamefont {I.}~\bibnamefont
			{Leuthäusser}},\ }\bibfield  {title} {{\selectlanguage {en}\bibinfo {title}
			{Statistical mechanics of {Eigen}'s evolution model}},\ }\href
	{https://doi.org/10.1007/BF01010413} {\bibfield  {journal} {\bibinfo
			{journal} {J. Stat. Phys.}\ }\textbf {\bibinfo {volume} {48}},\ \bibinfo
		{pages} {343} (\bibinfo {year} {1987})}\BibitemShut {NoStop}%
	\bibitem [{\citenamefont {Tkachenko}\ and\ \citenamefont
		{Maslov}(2015)}]{tkachenko_spontaneous_2015}%
	\BibitemOpen
	\bibfield  {author} {\bibinfo {author} {\bibfnamefont {A.~V.}\ \bibnamefont
			{Tkachenko}}\ and\ \bibinfo {author} {\bibfnamefont {S.}~\bibnamefont
			{Maslov}},\ }\bibfield  {title} {{\selectlanguage {en}\bibinfo {title}
			{Spontaneous emergence of autocatalytic information-coding polymers}},\
	}\href {https://doi.org/10.1063/1.4922545} {\bibfield  {journal} {\bibinfo
			{journal} {J. Chem. Phys.}\ }\textbf {\bibinfo {volume} {143}},\ \bibinfo
		{pages} {045102} (\bibinfo {year} {2015})}\BibitemShut {NoStop}%
	\bibitem [{\citenamefont {Matsubara}\ and\ \citenamefont
		{Kaneko}(2016)}]{matsubara_optimal_2016}%
	\BibitemOpen
	\bibfield  {author} {\bibinfo {author} {\bibfnamefont {Y.~J.}\ \bibnamefont
			{Matsubara}}\ and\ \bibinfo {author} {\bibfnamefont {K.}~\bibnamefont
			{Kaneko}},\ }\bibfield  {title} {{\selectlanguage {en}\bibinfo {title}
			{Optimal size for emergence of self-replicating polymer system}},\ }\href
	{https://doi.org/10.1103/PhysRevE.93.032503} {\bibfield  {journal} {\bibinfo
			{journal} {Phys. Rev. E}\ }\textbf {\bibinfo {volume} {93}},\ \bibinfo
		{pages} {032503} (\bibinfo {year} {2016})}\BibitemShut {NoStop}%
	\bibitem [{\citenamefont {Matsubara}\ and\ \citenamefont
		{Kaneko}(2018)}]{matsubara_kinetic_2018}%
	\BibitemOpen
	\bibfield  {author} {\bibinfo {author} {\bibfnamefont {Y.~J.}\ \bibnamefont
			{Matsubara}}\ and\ \bibinfo {author} {\bibfnamefont {K.}~\bibnamefont
			{Kaneko}},\ }\bibfield  {title} {{\selectlanguage {en}\bibinfo {title}
			{Kinetic {Selection} of {Template} {Polymer} with {Complex} {Sequences}}},\
	}\href {https://doi.org/10.1103/PhysRevLett.121.118101} {\bibfield  {journal}
		{\bibinfo  {journal} {Phys. Rev. Lett.}\ }\textbf {\bibinfo {volume} {121}},\
		\bibinfo {pages} {118101} (\bibinfo {year} {2018})}\BibitemShut {NoStop}%
	\bibitem [{\citenamefont {Tkachenko}\ and\ \citenamefont
		{Maslov}(2018)}]{tkachenko_onset_2018}%
	\BibitemOpen
	\bibfield  {author} {\bibinfo {author} {\bibfnamefont {A.~V.}\ \bibnamefont
			{Tkachenko}}\ and\ \bibinfo {author} {\bibfnamefont {S.}~\bibnamefont
			{Maslov}},\ }\bibfield  {title} {{\selectlanguage {en}\bibinfo {title} {Onset
				of natural selection in populations of autocatalytic heteropolymers}},\
	}\href {https://doi.org/10.1063/1.5048488} {\bibfield  {journal} {\bibinfo
			{journal} {J. Chem. Phys.}\ }\textbf {\bibinfo {volume} {149}},\ \bibinfo
		{pages} {134901} (\bibinfo {year} {2018})}\BibitemShut {NoStop}%
	\bibitem [{\citenamefont {Rosenberger}\ \emph {et~al.}(2021)\citenamefont
		{Rosenberger}, \citenamefont {Göppel}, \citenamefont {Kudella},
		\citenamefont {Braun}, \citenamefont {Gerland},\ and\ \citenamefont
		{Altaner}}]{rosenberger_self-assembly_2021}%
	\BibitemOpen
	\bibfield  {author} {\bibinfo {author} {\bibfnamefont {J.~H.}\ \bibnamefont
			{Rosenberger}}, \bibinfo {author} {\bibfnamefont {T.}~\bibnamefont
			{Göppel}}, \bibinfo {author} {\bibfnamefont {P.~W.}\ \bibnamefont
			{Kudella}}, \bibinfo {author} {\bibfnamefont {D.}~\bibnamefont {Braun}},
		\bibinfo {author} {\bibfnamefont {U.}~\bibnamefont {Gerland}},\ and\ \bibinfo
		{author} {\bibfnamefont {B.}~\bibnamefont {Altaner}},\ }\bibfield  {title}
	{{\selectlanguage {en}\bibinfo {title} {Self-{Assembly} of {Informational}
				{Polymers} by {Templated} {Ligation}}},\ }\href
	{https://doi.org/10.1103/PhysRevX.11.031055} {\bibfield  {journal} {\bibinfo
			{journal} {Phys. Rev. X}\ }\textbf {\bibinfo {volume} {11}},\ \bibinfo
		{pages} {031055} (\bibinfo {year} {2021})}\BibitemShut {NoStop}%
	\bibitem [{\citenamefont {Kudella}\ \emph {et~al.}(2021)\citenamefont
		{Kudella}, \citenamefont {Tkachenko}, \citenamefont {Salditt}, \citenamefont
		{Maslov},\ and\ \citenamefont {Braun}}]{kudella_structured_2021}%
	\BibitemOpen
	\bibfield  {author} {\bibinfo {author} {\bibfnamefont {P.~W.}\ \bibnamefont
			{Kudella}}, \bibinfo {author} {\bibfnamefont {A.~V.}\ \bibnamefont
			{Tkachenko}}, \bibinfo {author} {\bibfnamefont {A.}~\bibnamefont {Salditt}},
		\bibinfo {author} {\bibfnamefont {S.}~\bibnamefont {Maslov}},\ and\ \bibinfo
		{author} {\bibfnamefont {D.}~\bibnamefont {Braun}},\ }\bibfield  {title}
	{{\selectlanguage {en}\bibinfo {title} {Structured sequences emerge from
				random pool when replicated by templated ligation}},\ }\href
	{https://doi.org/10.1073/pnas.2018830118} {\bibfield  {journal} {\bibinfo
			{journal} {Proc. Natl. Acad. Sci. U.S.A.}\ }\textbf {\bibinfo {volume}
			{118}},\ \bibinfo {pages} {e2018830118} (\bibinfo {year} {2021})}\BibitemShut
	{NoStop}%
	\bibitem [{\citenamefont {Göppel}\ \emph {et~al.}(2022)\citenamefont
		{Göppel}, \citenamefont {Rosenberger}, \citenamefont {Altaner},\ and\
		\citenamefont {Gerland}}]{goppel_thermodynamic_2022}%
	\BibitemOpen
	\bibfield  {author} {\bibinfo {author} {\bibfnamefont {T.}~\bibnamefont
			{Göppel}}, \bibinfo {author} {\bibfnamefont {J.~H.}\ \bibnamefont
			{Rosenberger}}, \bibinfo {author} {\bibfnamefont {B.}~\bibnamefont
			{Altaner}},\ and\ \bibinfo {author} {\bibfnamefont {U.}~\bibnamefont
			{Gerland}},\ }\bibfield  {title} {{\selectlanguage {en}\bibinfo {title}
			{Thermodynamic and {Kinetic} {Sequence} {Selection} in {Enzyme}-{Free}
				{Polymer} {Self}-{Assembly} inside a {Non}-equilibrium {RNA} {Reactor}}},\
	}\href {https://doi.org/10.3390/life12040567} {\bibfield  {journal} {\bibinfo
			{journal} {Life}\ }\textbf {\bibinfo {volume} {12}},\ \bibinfo {pages} {567}
		(\bibinfo {year} {2022})}\BibitemShut {NoStop}%
	\bibitem [{\citenamefont {Losick}(1972)}]{losick_vitro_1972}%
	\BibitemOpen
	\bibfield  {author} {\bibinfo {author} {\bibfnamefont {R.}~\bibnamefont
			{Losick}},\ }\bibfield  {title} {\bibinfo {title} {In {Vitro}
			{Transcription}},\ }\href
	{https://doi.org/https://doi.org/10.1146/annurev.bi.41.070172.002205}
	{\bibfield  {journal} {\bibinfo  {journal} {Annu. Rev. Biochem.}\ }\textbf
		{\bibinfo {volume} {41}},\ \bibinfo {pages} {409} (\bibinfo {year}
		{1972})}\BibitemShut {NoStop}%
	\bibitem [{\citenamefont {Mullis}\ \emph {et~al.}(1986)\citenamefont {Mullis},
		\citenamefont {Faloona}, \citenamefont {Scharf}, \citenamefont {Saiki},
		\citenamefont {Horn},\ and\ \citenamefont {Elrich}}]{mullis_specific_1986}%
	\BibitemOpen
	\bibfield  {author} {\bibinfo {author} {\bibfnamefont {K.}~\bibnamefont
			{Mullis}}, \bibinfo {author} {\bibfnamefont {F.}~\bibnamefont {Faloona}},
		\bibinfo {author} {\bibfnamefont {S.}~\bibnamefont {Scharf}}, \bibinfo
		{author} {\bibfnamefont {R.}~\bibnamefont {Saiki}}, \bibinfo {author}
		{\bibfnamefont {G.}~\bibnamefont {Horn}},\ and\ \bibinfo {author}
		{\bibfnamefont {H.}~\bibnamefont {Elrich}},\ }\bibfield  {title} {\bibinfo
		{title} {Specific {Enzymatic} {Amplification} of {DNA} {In} {Vitro}: {The}
			{Polymerase} {Chain} {Reaction}},\ }\href
	{https://doi.org/10.1101/SQB.1986.051.01.032} {\bibfield  {journal} {\bibinfo
			{journal} {Cold Spring Harb. Symp. Quant. Biol.}\ }\textbf {\bibinfo
			{volume} {51}},\ \bibinfo {pages} {263} (\bibinfo {year} {1986})}\BibitemShut
	{NoStop}%
\end{thebibliography}

%

\end{document}


\title{Supplemental Information for \\ `Non Equilibrium Transitions in a Template Copying Ensemble'}

\author{Arthur Genthon}
\author{Carl D. Modes}
\author{Frank Jülicher}
\author{Stephan W. Grill}

\maketitle

\renewcommand{\theequation}{S\arabic{equation}}
\setcounter{equation}{0}
\renewcommand{\thefigure}{S\arabic{figure}}
\setcounter{figure}{0}

\tableofcontents

\section{Statistics of errors}

\subsection{Solving the master equation}

We recast the master equation (eq. 3 in the main text) as an partial differential equation for the generating function
\begin{equation}
    G(z,t)=\sum_{N=0}^{\infty} z^N p(N,t)
\end{equation}
by multiplying it by $z^N$ and summing over $N$:
\begin{equation}
    \partial_t G(z,t) = -k_d (z-1) \partial_z G(z,t) +k_a (z-1) G(z,t) \,.
\end{equation}
This equation is solved with the method of characteristics by choosing the parametrization $\hat{G}(x)=G(z(x),t(x))$ with $t(x)=x$ and $dz/dx = k_d (z-1)$ such that the ordinary differential equation
\begin{equation}
   \frac{d\hat{G}}{dx}=k_a (z(x)-1) \hat{G}(x) 
\end{equation}
is solvable and gives the solution
\begin{equation}
    \hat{G}(x) =\hat{G}(0)  \exp \lbk \int_0^x k_a (z(x')-1) dx' \rbk \,,
\end{equation}
where $\hat{G}(0)=G(z(0),0)\equiv G_0(z(0))$ is the initial condition. 

Integrating $dz/dx$ gives $z(x)=(z(0)-1) \exp(k_d x) +1$, so that the solution reads:
\begin{align}
    G(z,t) &=G_0\lp1+(z-1) e^{-k_d t} \rp e^{(z-1) \lambda(t)} \\
    \lambda(t) &= \frac{k_a}{k_d} \lp 1 - e^{-k_d t} \rp \,.
\end{align}

We start at time $t=0$ with no copy sequences in our system, $p(N,t=0)=\delta_N$, which implies $G_0=1$. Finally, the solution is a Poisson distribution at all times, with rate $\lambda(t)$:
\begin{equation}
    \label{eq:poisson}
    p(N,t)=\frac{\lambda(t)^N}{N!} e^{-\lambda(t)} \,.
\end{equation}
For any length $L$, $k_d$ is finite and the solution converges towards the steady-state Poisson  distribution with rate $k_a/k_d$.

\subsection{Distribution for the number of copies with $q$ errors}

We now derive the distribution $Q(N_q,t)$ for the number $N_q$ of polymers with $q$ errors when compared to the template at time $t$. There are $\Omega_q={L \choose q} (m-1)^q$ sequences with $q$ errors, which we label $1 \leq i_1 < ... < i_{\Omega_q} \leq m^L$, associated with numbers of copies $N_{i_1}$, ..., $N_{i_{\Omega_q}}$. The distribution thus reads:
\begin{equation}
   Q(N_q,t)=\sum_{N_{i_1} + ... + N_{i_{\Omega_q}}=N_q} p(N_{i_1}, ..., N_{i_{\Omega_q}},t) \,.
\end{equation}
Since sequences are independent, the joint distribution of copy numbers is the product of the marginal distributions given by \cref{eq:poisson}, which share the same rate $\lambda_q(t)$. Therefore, using the multinomial theorem, we get that the distribution $Q$ is also Poissonian with rate $\lambda_q(t) \Omega_q$:
\begin{align}
    Q(N_q,t) &= \lambda_q(t)^{N_q}e^{-\lambda_q(t)\Omega_q}\sum_{N_{i_1} + ... + N_{i_{\Omega_q}}=N_q} \prod_{j=1}^{\Omega_q} (N_{i_j}!)^{-1}\\
    &= \frac{(\lambda_q(t)\Omega_q)^{N_q}}{N_q!} e^{-\lambda_q(t)\Omega_q} \,.
\end{align}
Finally, the average number of polymers with error fraction $x=q/L$ is equal to 
\begin{equation}
    \langle N_x(t) \rangle =\lambda_{xL}(t)\Omega_{xL} \,.
\end{equation}

\section{Phase diagram}
\label{app_phas_diag}

\subsection{Kinetic rates parametrization}

The micro-reversibility condition constrains the ratio of the forward to backward rates for each process: $k_j^+/k_j^- = e^{-(\Delta \mu_r - \Delta \mu_F)L}$ and $k_r^+/k_r^- = e^{-\Delta \mu_r L}$, which leaves some freedom in the individual definitions of these rates. We consider the following general parametrization: 
\begin{align}
	\label{eq:kj+}
	k_j^+ &=k_j e^{-((1+\alpha)\Delta \mu_r - (1+\gamma) \Delta \mu_F)L} \\
	\label{eq:kj-}
	k_j^- &=k_j e^{-(\alpha\Delta \mu_r - \gamma \Delta \mu_F)L} \\
	\label{eq:kr+}
	k_r^+ &=k_r e^{-(1+\beta)\Delta \mu_r L} \\
	\label{eq:kr-}
	k_r^- &=k_r e^{-\beta\Delta \mu_r L} \,,
\end{align}
where the energetic biases $\Delta \mu_r$ and $\Delta \mu_F$ are arbitrarily split between the forward and backward rates with parameters $\alpha, \beta, \gamma \in \mathbb{R}$. 
The case treated in the first part of the main text corresponds to the choice $\alpha=\beta=\gamma=0$, and the introduction of the fuel-dependent energy barrier in the last section of the main text corresponds to the choice $\alpha=\beta=0$ and $\gamma \neq 0$. 
In this section we treat the most general case.

\subsection{General phase diagram}

We now investigate which sequences are dominant in the steady-state population in the large $L$ limit. To do so, we define the monomer error fraction, $x=q/L$, and seek the maxima of the expected number $\langle N_x \rangle=\lambda_{xL} \Omega_{xL}$ of sequences with an error fraction $x$. 

Since $\langle N_x \rangle$ grows exponentially with $L$, it is convenient to instead work with
\begin{align}
    f(x) &=-\frac{1}{L} \ln \langle N_x \rangle \\
    &= -\frac{1}{L} \ln \lbk {L \choose xL} (m-1)^{xL} \ e^{ -\Delta \mu_r L} \ \frac{ k_0 e^{(-ax - \alpha\Delta \mu_r + (1+\gamma) \Delta \mu_F )L} + k_r e^{-\beta \Delta \mu_r L} }{ k_0 e^{(-ax - \alpha\Delta \mu_r + \gamma \Delta \mu_F )L} + k_r e^{-\beta \Delta \mu_r L} } \rbk
\end{align}
We use Stirling formula, $n! = \sqrt{2\pi n} (n/e)^n (1 + O(1/n))$ for large $n$:
\begin{equation}
    {L \choose xL} = (2\pi L x(1-x))^{-1/2} x^{-xL} (1-x)^{-(1-x)L} \lp 1+O\lp \frac{1}{L}\rp\rp \,.
\end{equation}
We simplify the denominator and numerator of the fraction as 
\begin{align}
\label{eq_ln_bw}
    \frac{1}{L}\ln \lp k_0 e^{(-ax - \alpha\Delta \mu_r + \gamma \Delta \mu_F )L} + k_r e^{-\beta \Delta \mu_r L} \rp = & \lp -ax - \alpha\Delta \mu_r + \gamma \Delta \mu_F  + \frac{\ln k_0}{L} \rp \theta \lp x_m - x \rp \nn \\
    & + \lp -\beta \Delta \mu_r + \frac{\ln k_r}{L} \rp \theta \lp x -x_m \rp +o\lp \frac{1}{L} \rp \\
    \label{eq_ln_fw}
    \frac{1}{L}\ln \lp k_0 e^{(-ax - \alpha\Delta \mu_r + (1+\gamma) \Delta \mu_F )L} + k_r e^{-\beta \Delta \mu_r L} \rp = & \lp -ax - \alpha\Delta \mu_r + (1+\gamma) \Delta \mu_F + \frac{\ln k_0}{L} \rp \theta \lp x_M - x \rp \nn \\
    & + \lp -\beta \Delta \mu_r + \frac{\ln k_r}{L} \rp \theta \lp x -x_M \rp +o\lp \frac{1}{L} \rp
\end{align}
with $\theta$ the Heaviside function and the threshold error fractions
\begin{align}
x_m &= \frac{(\beta-\alpha) \Delta \mu_r +\gamma \Delta \mu_F }{a} \\
x_M &= \frac{(\beta-\alpha) \Delta \mu_r + (1+\gamma) \Delta \mu_F  }{a} \,,
\end{align}
which are such that $x_M-x_m=\Delta \mu_F  /a >0$.

Combining these results we obtain
\begin{align}
\label{eq_f_cases}
f(x) 
= g(x) + \frac{1}{2L} \ln \lp 2 \pi L x (1-x) \rp + \Delta \mu_r + 
    \begin{cases}
    - \Delta \mu_F + o\lp \frac{1}{L} \rp & \text{if } x < x_m \\
    ax + (\alpha - \beta) \Delta \mu_r - (1+\gamma) \Delta \mu_F + \frac{\ln(k_r/k_0)}{L}+ o\lp \frac{1}{L} \rp & \text{if } x_m < x < x_M \\
    o\lp \frac{1}{L} \rp & \text{if } x_M < x
    \end{cases}
\end{align}
where we defined the function
\begin{equation}
    g(x)=x\ln\lp \frac{x}{m-1} \rp +(1-x)\ln(1-x) \,.
\end{equation}

We seek the minima of $f$ at first order in $1/L$: 
\begin{equation}
    x=x^{(0)}+\frac{x^{(1)}}{L} + o\lp \frac{1}{L} \rp. 
\end{equation}
At zero-th order, the minima of $g(x)$ and of $h(x)=g(x)+ax$ are respectively: 
\begin{align}
    x_r^{(0)} &= \frac{m-1}{m} \\
    x_a^{(0)} &= \frac{1}{1 + e^{a}/(m-1)} \,,
\end{align}
with $g(x_r^{(0)})=-\ln m$ and $h(x_a^{(0)})=-\ln \lp 1+(m-1) e^{-a} \rp$.

The first order corrections are obtained by injecting $x=x^{(0)}+x^{(1)}/L$ in $f'(x)=0$, which leads to 
\begin{align}
    x^{(1)} &= - \frac{\lp \ln \lp  x (1-x) \rp \rp'}{2 g''(x)} \Bigg\rvert_{x=x^{(0)}} \\
    & = \frac{2 x^{(0)} - 1}{2} \,,
\end{align}
both for the accurate and random copies, since $g''(x) = (g(x) + ax)''$.

\begin{figure}
    \includegraphics[width=0.32\linewidth]{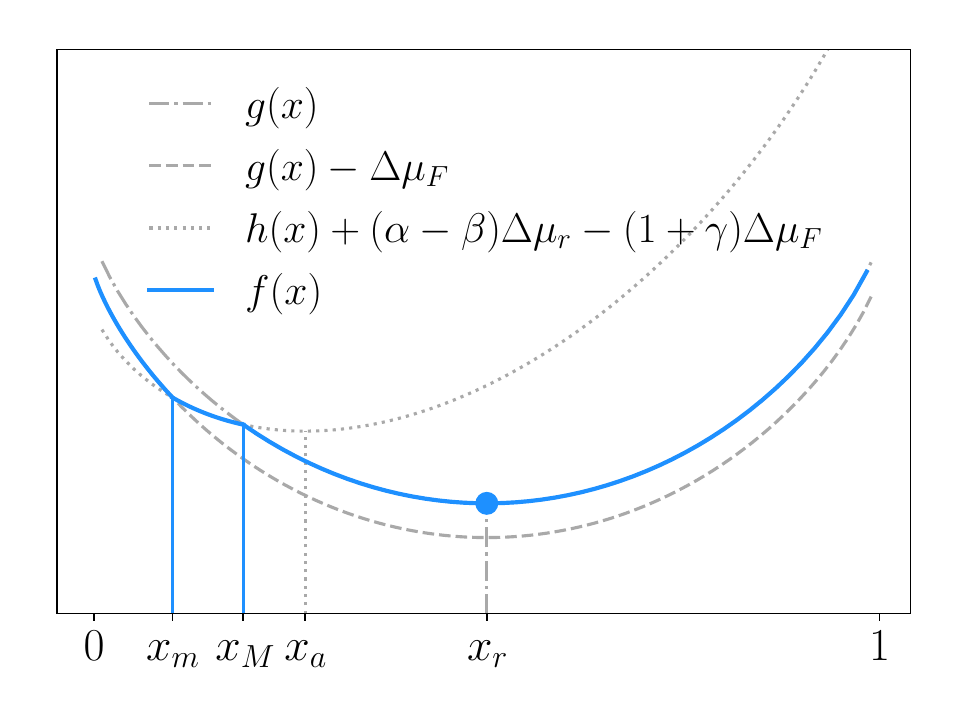}
    \includegraphics[width=0.32\linewidth]{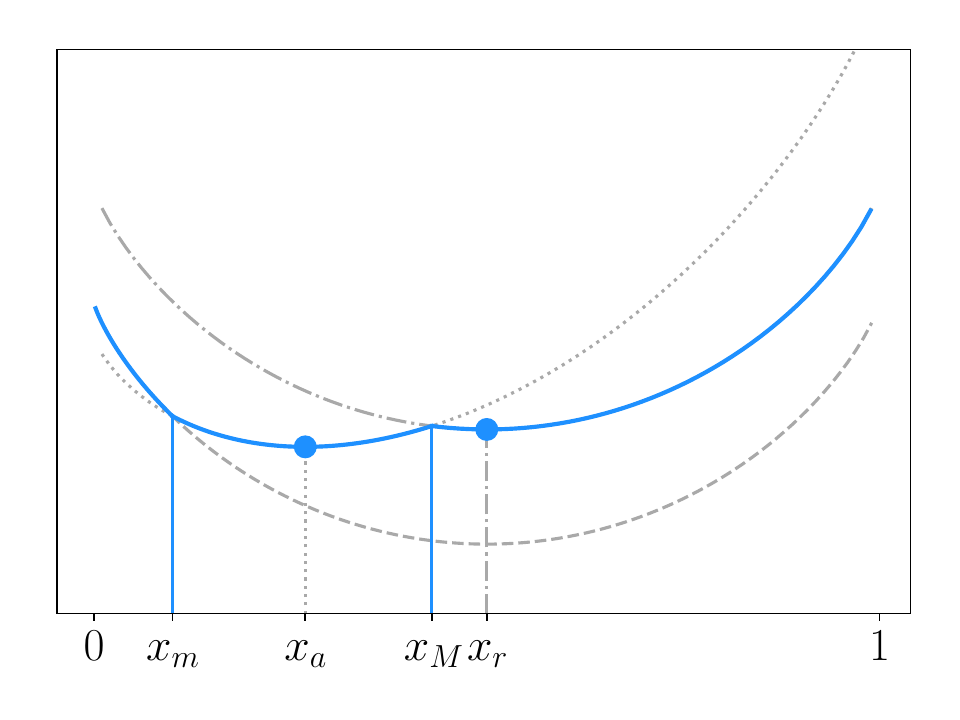}
    \includegraphics[width=0.32\linewidth]{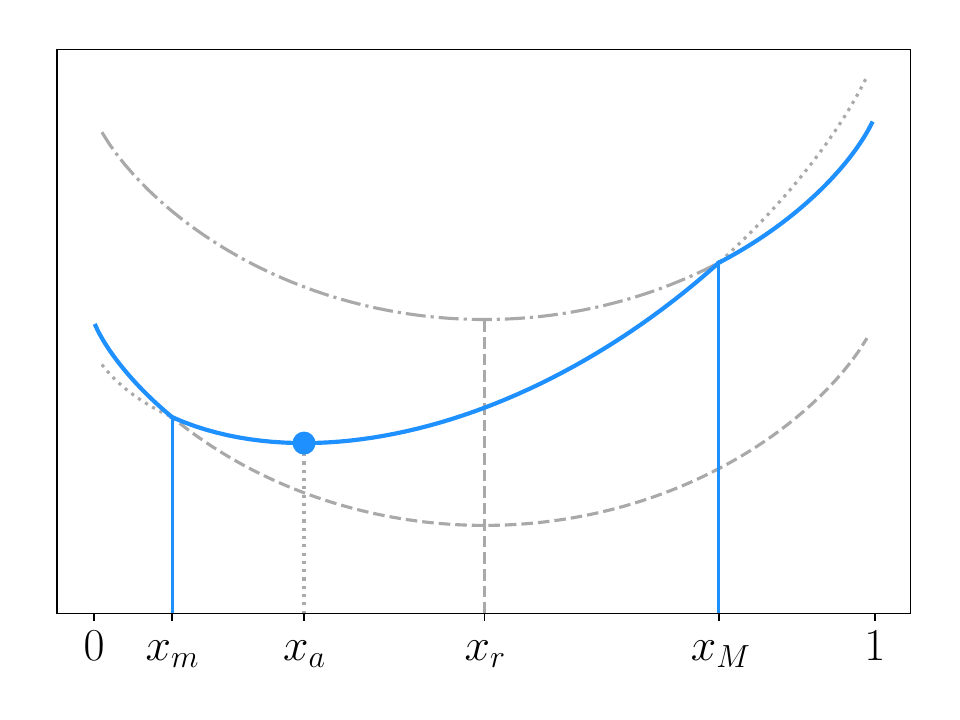}
    \includegraphics[width=0.32\linewidth]{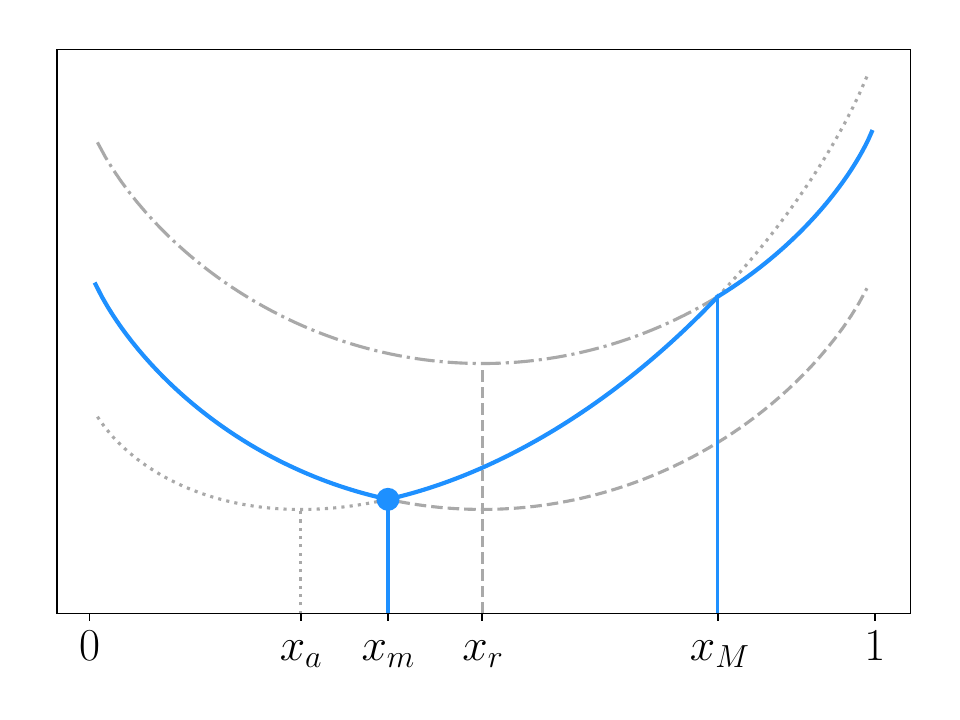}
    \includegraphics[width=0.32\linewidth]{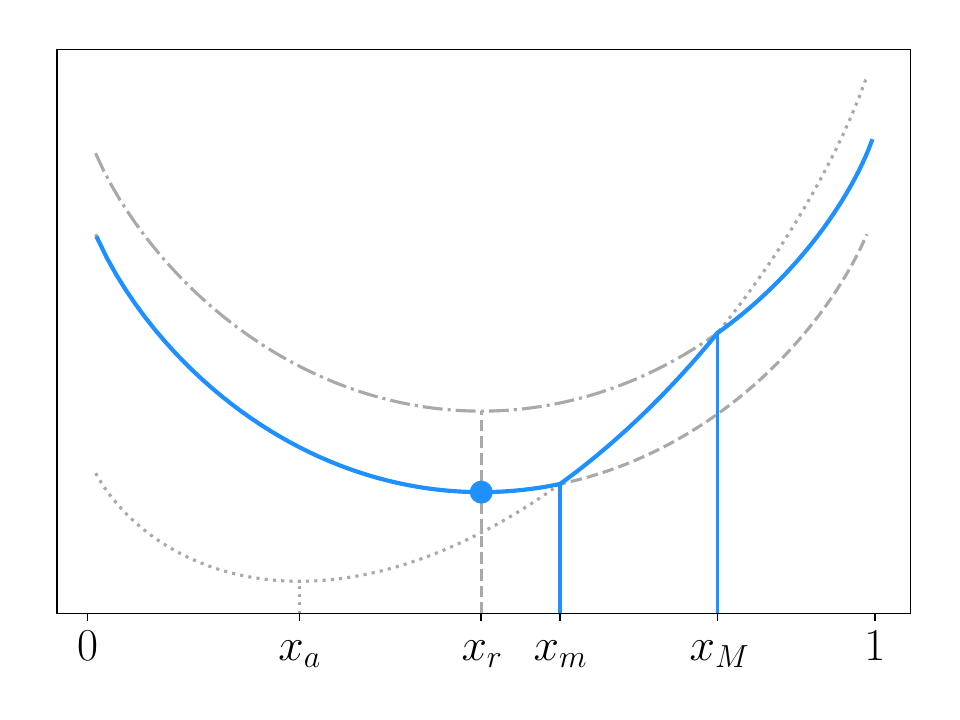}    
    \includegraphics[width=0.32\linewidth]{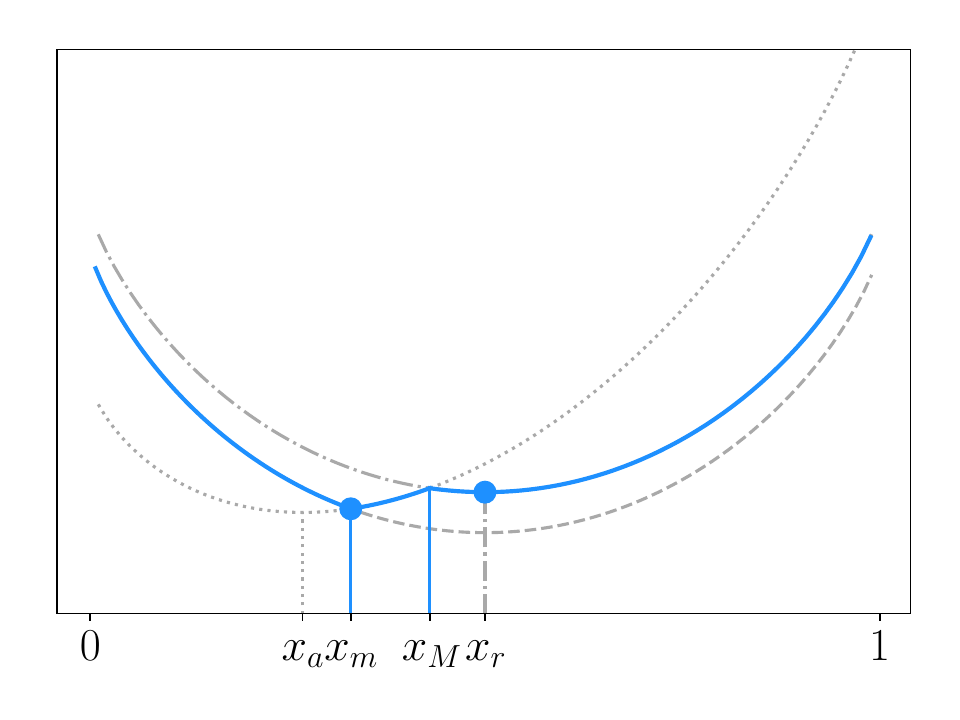}
    \caption{Plots of function $f(x) =-\ln \langle N_x \rangle$/L, shown in blue, versus error fraction $x$, in the 6 cases defined in the text. The grey dotted, dashed and dash-dotted curves are the three functions that define $f$ in a piece-wise manner in \cref{eq_f_cases}. The minima of function $f$ are indicated with blue circles.}
    \label{fig:f_cases}
\end{figure}

Whether or not these minima $x_r$ and $x_a$ are reached by $f$ depends on their relative positions with the threshold error fractions $x_m$ and $x_M$. For example, if $x_a \in [x_m,x_M]$, where $f$ is equal to $h$ (+ corrections), then $f$ reaches a local minimum at $x_a$, while if $x_a \notin [x_m,x_M]$, then $f$ has no local minimum on $[x_m,x_M]$.
There are 6 cases (because $x_a < x_r$ and $x_m < x_M$), listed below, and represented on \cref{fig:f_cases} where the dots indicate the minima of $f$.
\begin{enumerate}    
    \item $x_m < x_M < x_a < x_r$: Global minimum $x_r$
    \item $x_m < x_a < x_M < x_r$: Local minima $(x_a, x_r)$
    \item $x_m < x_a < x_r < x_M$: Global minimum $x_a$
    \item $x_a < x_m < x_r < x_M$: Global minimum $x_m$
    \item $x_a < x_r < x_m < x_M$: Global minimum $x_r$
    \item $x_a < x_m < x_M < x_r$: Local minima $(x_m, x_r)$
\end{enumerate} 
In the end, $f$ has either one or two minima, corresponding to one or two peaks in the population $\langle N_x \rangle$. Note that since $x_m$ is threshold error fraction for the definition of the piece-wise function $f$, it is a local minimum of $f$ only when $f$ is decreasing on $[0,x_m]$ and increasing on $[x_m,x_M]$, i.e. when $x_a < x_m < x_r$. 

In cases 2 and 6, where $\langle N_x \rangle$ is bimodal, the relative height of the two peaks is given by the difference in the values of $f$ at the two error fractions. For example, in case 2 we have
\begin{align}
    f(x_r) &=  -\ln(m)+ \frac{1}{2L} \ln \lp 2 \pi L x_r^{(0)} (1-x_r^{(0)}) \rp + \Delta \mu_r + o\lp \frac{1}{L} \rp\\
    f(x_a) &= -\ln \lp 1+(m-1) e^{-a} \rp + \frac{1}{2L} \ln \lp 2 \pi L x_a^{(0)} (1-x_a^{(0)}) \rp +  (1+\alpha - \beta) \Delta \mu_r - (1+\gamma) \Delta \mu_F + \frac{\ln(k_r/k_0)}{L}+ o\lp \frac{1}{L} \rp \,.
\end{align}
The two populations are thus equal, $\langle N_{x_r} \rangle = \langle N_{x_a} \rangle$ (peaks of the same height), when $\Delta \mu_F=\Delta \mu_F^*$ with
\begin{equation}
	\label{eq_delta_mu_f_*}
        \Delta \mu_F^*= \frac{1}{1+\gamma} \lbk \ln \lp \frac{m}{1+(m-1) e^{-a}} \rp - (\beta-\alpha) \Delta \mu_r + \frac{1}{L} \ln \lp \frac{k_r}{k_0}\sqrt{\frac{x_a^{(0)}(1-x_a^{(0)})}{x_r^{(0)}(1-x_r^{(0)})}}\rp + o\lp\frac{1}{L}\rp \rbk\,.
\end{equation}
When $\alpha=\beta=\gamma=0$, this result gives back eq. (7) from the main text.

\begin{figure}
    \centering
    \includegraphics[width=0.7\linewidth]{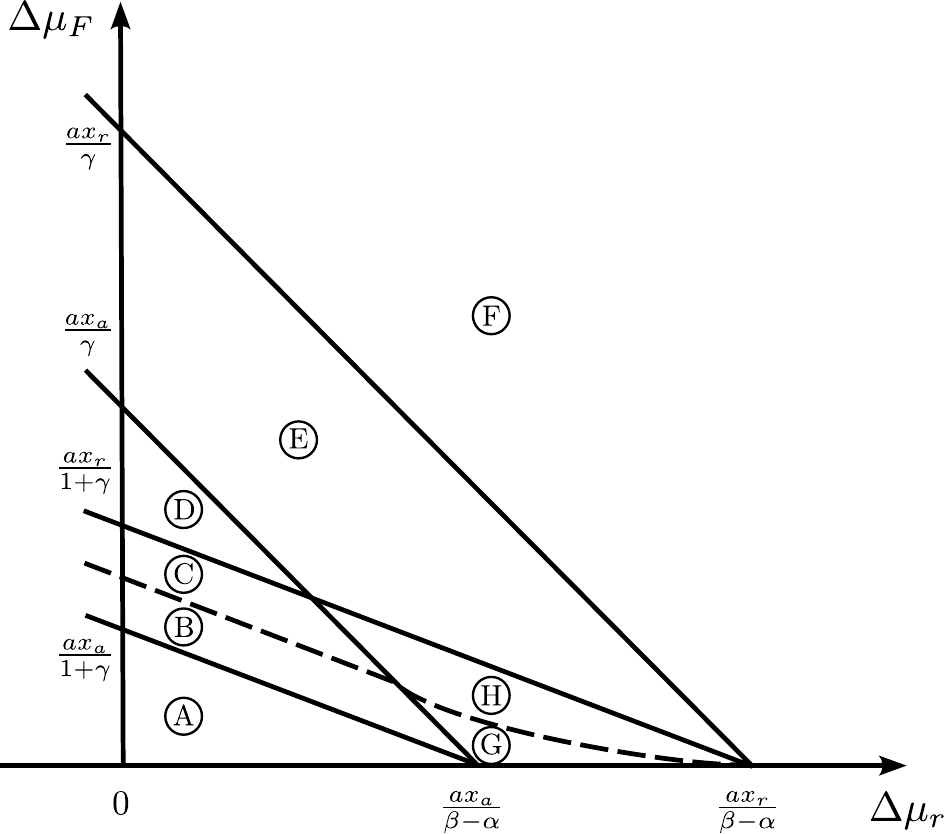}
    \caption{Phase diagram for a fixed value of specificity $a$, in the case $\alpha < \beta$ and $\gamma >0$. The regions A to H are characterized by the existence of one or two peaks of the average number of copies $\langle N_x \rangle$ at error fractions specified in the text. }
    \label{fig:full_phase_diag}
\end{figure}

We show on \cref{fig:full_phase_diag} an example of general phase diagram. 
The regions A to H correspond to the cases 1 to 6 in the following manner: the regions A, D, E and F of unimodal population $\langle N_x \rangle$ correspond to the cases 1, 3, 4 and 5 respectively. The regions (B,C) and (G,H) of bimodal population $\langle N_x \rangle$ correspond respectively to the two sub-regions of cases 2 and 6, defined by the dominance of one or other of the two error fractions, and are delimited by dashed curves of equal populations.  The inequalities between $x_m$, $x_M$, $x_a$ and $x_r$ which define the 6 cases give the equations of the phase borders.
To plot this phase diagram, we chose $\alpha < \beta$ and $\gamma>0$, but similar diagrams are obtained for other values of the parameters. 

In the simple case treated first in the main text, namely when $\alpha=\beta=\gamma=0$, then $x_m=0$, and thus cases 4, 5 and 6 are not possible and regions E to H are not accessible. 
More generally, when $\alpha=\beta$ all the phase borders become horizontal, and regions G and H are not accessible. 
In this case, the boundary $\Delta \mu_F^*$ between random and accurate copies (dashed line between B and C) is independent of $\Delta \mu_r$, i.e. the accuracy of copies is independent of the intrinsic free energy difference between polymers and monomers. This is because accuracy results from the competition between the random and templated processes, and $\Delta \mu_r$ plays no role in this competition when $\alpha=\beta$. However, the absolute values of the kinetics rates, and thus $\Delta \mu_r$, control the population size, as shown in \cref{sec_pop_size}.

\subsection{Fuel-dependent energy barrier}

When $\alpha=\beta=\gamma=0$, the forward rates $k_j^+$ and $k_r^+$ depend on energetics but the backward rates $k_j^-$ and $k_r^-$ do not. In this case, the rate of templated disassembly $k_j^-=k_0 e^{-axL}$ vanishes for large $L$ while spontaneous disassembly is constant $k_r^-=k_r$. Hence, disassembly in the large $L$ limit is dominated by the spontaneous process. Introducing a fuel-dependent energy barrier with $\gamma >0$ allows templated disassembly to compete with spontaneous disassembly for some values of the error faction $x$. We further consider the case $\alpha=\beta$, where the energy difference $\Delta \mu_r$ between monomers and polymer has the same effect on the assembly rates of the templated and the spontaneous processes, such that the energy barriers in each case have the same $L$ dependence. 

In this section, we detail the phases A to H of the phase diagram shown in fig 4 in the case of a fuel-dependent energy barrier which speeds up the templated process, namely with $\gamma>0$. While in the main text $\alpha=\beta=0$ is implicitly assumed, note that the value of $\alpha=\beta$ has no impact on the phase diagram, as explained in the previous section.

In fig 4a, regions A to D are characterized by the same error statistics as the four regions of fig 2a. Compared to the phase diagram for $\alpha=\beta=\gamma=0$ shown in fig 2a, the boundaries of regions A to D are scaled by a factor $1+\gamma$, but the statistics of error fractions in these regions remain the same. In addition, fig 4a comprises four new regions E to H. 
In the new region E the distribution of copying errors in the large $L$ limit is unimodal with a single peak at error fraction
$x_m=\gamma\Delta \mu_F/a$.
Region E is separated from region F by $\Delta \mu_F = a x_r / \gamma$, from region H by $\Delta \mu_F = a x_r / (1+\gamma)$, and from region D by $\Delta \mu_F = a x_a / \gamma$. As $\Delta \mu_F$ is increased, $x_m$  increases linearly from $x_a$ at the boundary between regions D and E to $x_r$ at the boundary between regions E and F. We thus call $x_m$ the intermediate error fraction.
In region F, where $\Delta \mu_F > a x_r / \gamma$, assembly and disassembly are both governed by the templated process for error fractions $x \leq x_r$. Since disassembly is now also sequence-dependent, accurate copies are disassembled faster than inaccurate ones, and the steady-state generated by the templated process becomes sequence-independent with dominating error fraction $x_r$. 
Regions G and H are regions of phase coexistence between random copies with error fraction $x_r$ and copies with intermediate error fraction $x_m$. These regions are separated from region E by $\Delta \mu_F = a x_r / (1+\gamma)$ and from regions B and C by $\Delta \mu_F = a x_a / \gamma$. Regions G and H are separated from each other by the non-analytic curve representing the equality between the copy numbers of both phases, shown in light blue in fig 4a. 

\sloppy We show in fig 4b the phase diagram in the large $L$ limit for a fixed value of specificity $a$ such that $\ln \lp m/(1+(m-1)e^{-a}) \rp /(1+\gamma) < a x_a/\gamma$. 
Similar to the example shown in fig 2b, for small $\Delta \mu_F$, the system generates random copies. The transition from random to accurate copies occurs at the threshold $\Delta \mu_F^*$ given by \cref{eq_delta_mu_f_*}, which is reduced by a factor $1+\gamma$ relative to the value given by eq. (7) in the main text.
Further increasing $\Delta \mu_F$ leads to a continuous increase in the average error fraction until the system re-enters a regime of random copying. 
This is because disassembly becomes dominated by the templated process, which is sequence-specific. Therefore, if accurate copies are assembled faster than inaccurate copies via templated assembly, they are also disassembled faster via templated disassembly, which results in a random distribution. From the thermodynamic point of view, when both assembly and disassembly are dominated by only process, be it spontaneous or templated, then this process is in detailed balance in the steady-state limit, and thus can only produce the equilibrium distribution which is the distribution centred around random copies since all sequences have the same energy. 
There is thus a range of values $\Delta \mu_F^{*} < \Delta \mu_F < a x_a / \gamma$ where a maximum accuracy at error fraction $x_a$ is achieved. 
In the large $L$ limit, the number of copies can vanish, which is indicated by the white region in fig 4b. Extinction conditions are given explicitly in \cref{sec_pop_size}.

\section{Phase transition}

\begin{figure}
    \centering
    \includegraphics[width=0.6\linewidth]{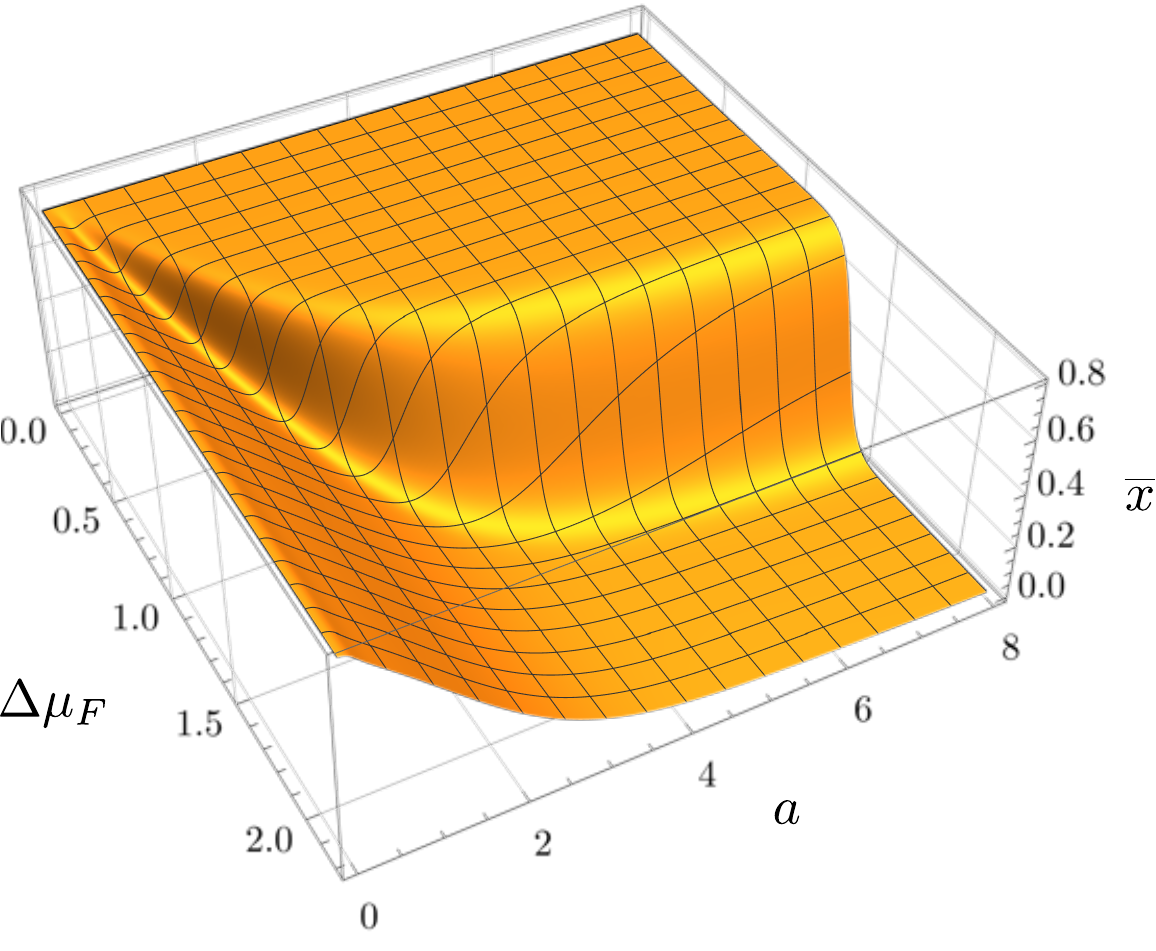}
    \caption{Average error fraction $\overline{x}$ vs energy drive $\Delta \mu_F$ and specificity $a$ for $m=4$, $k_0=1$, $k_r=0.1$, $\Delta \mu_r=0.5$ and $L=25$.}
    \label{fig:x_vs_DmuF_vs_a}
\end{figure}

We show in \cref{fig:x_vs_DmuF_vs_a} the 3-dimensional surface plot of the average error fraction $\overline{x}$ as a function of the energy drive $\Delta \mu_F$ and of the specificity $a$, for $L=25$ and for values of parameters equal to those of figure 3 in the main text. When increasing $\Delta \mu_F$, we observe the transition from random copies with error fraction $x_r$ to accurate copies with error fraction $x_a$ at threshold value $\Delta \mu_F^*(a)$, for all values of $a$. On the other hand, when increasing $a$, the transition from accurate copies to random copies occurs at threshold value $a^*(\Delta \mu_F)$ only for values of $\Delta \mu_F<\Delta \mu_F^{*,\infty}=\ln m +O(L^{-1})$. 

\section{Population size}
\label{sec_pop_size}

In this section we investigate the conditions under which the numbers of copies with error fractions $x_r$, $x_a$ and $x_m$ vanish in the large $L$ limit. Regions of vanishing population are represented in white in Figures 2b and 4b in the main text. 

In region A, the average number of copies $\langle N_x \rangle$ is characterized by a single peak around $x=x_r$. This region corresponds to case 1 in \cref{app_phas_diag} where $x_M < x_r$, so that in the large L limit \cref{eq_f_cases} reads 
\begin{equation}
    f(x_r)= - \ln m + \Delta \mu_r \,.
\end{equation}
The population of random copies therefore goes extinct if $\Delta \mu_r>\ln m$. 

In region F, the average number of copies $\langle N_x \rangle$ is also characterized by a single peak around $x=x_r$. This region corresponds to case 5 in \cref{app_phas_diag} where $x_r < x_m$, so that in the large L limit \cref{eq_f_cases} reads 
\begin{equation}
    f(x_r)= - \ln m + \Delta \mu_r - \Delta \mu_F \,.
\end{equation}
The population of random copies therefore goes extinct if $\Delta \mu_F < \Delta \mu_r - \ln m$. The line separating the phases of no population in white and random copies in red for $\Delta \mu_F>a x_r/\gamma$ in Fig and 4b is thus of slope 1. 

The reason why these two conditions are different even though the accuracy $x_r$ of the copies is the same in regions A and F is because in region A, both assembly and disassembly are controlled by the spontaneous process while in region F, both are controlled by the templated process. In both regions A and F, the steady-state generated by these processes is sequence-independent with dominating error fraction $x_r$, but the energies involved are different. The fuel burnt in the templated reaction allows for non-vanishing populations of copies for values of $\Delta \mu_r$ larger than $\ln m$. 

In region D, the average number of copies $\langle N_x \rangle$ is characterized by a single peak around $x=x_a$. This region corresponds to case 3 in \cref{app_phas_diag} where $x_m < x_a < x_M$, so that in the large L limit \cref{eq_f_cases} reads 
\begin{equation}
    f(x_a)= - \ln (1+ (m-1) e^{-a}) + (1+\alpha-\beta) \Delta \mu_r - (1+ \gamma) \Delta \mu_F \,.
\end{equation}
The population of accurate copies therefore goes extinct if $\Delta \mu_F < ((1+\alpha-\beta) \Delta \mu_r - \ln (1+ (m-1) e^{-a}))/(1+\gamma)$. In Fig 4b, plot in the case $\alpha=\beta$ and $\gamma>0$, the line separating the phases of no population in white and accurate copies in green is thus of slope $1/(1+\gamma)$, and in Fig 2b, plot in the case $\alpha=\beta=\gamma=0$, the same line has a slope $1$.

In region E, the average number of copies $\langle N_x \rangle$ is characterized by a single peak around $x=x_m$. This region corresponds to case 4 in \cref{app_phas_diag}, so that in the large L limit \cref{eq_f_cases} reads 
\begin{equation}
    f(x_m)= g(x_m) + \Delta \mu_r - \Delta \mu_F \,.
\end{equation}
The population of intermediate copies therefore goes extinct if $\Delta \mu_F < \Delta \mu_r + g(x_m)$. This condition is non-analytical in $\Delta \mu_F$, and is represented by the curve separating the phases of no population in white and intermediate copies in yellow in Fig 4b. One straightforwardly shows that there is no discontinuity in the derivative between this curve and the straight lines in the neighboring regions of accurate and intermediate copies.  

In regions of phase coexistence, B-C and G-H, the conditions for the extinction of the population associated with each error fraction are the same as the ones given above. For example, in regions B and C, the average number of copies $\langle N_x \rangle$ is characterized by two peaks around $x=x_r$ and $x=x_a$. These two regions correspond to case 2 in \cref{app_phas_diag}, where $x_m < x_a < x_M < x_r$. Thus, the population of random copies goes extinct if $\Delta \mu_r>\ln m$, like in region A where $x_M < x_r$ as well, and the population of accurate copies goes extinct if $\Delta \mu_F < ((1+\alpha-\beta)\Delta \mu_r - \ln (1+ (m-1) e^{-a}))/(1+\gamma)$ like in region D where $x_m < x_a < x_M$ as well.

\section{Non-equilibrium current}

We compute here the net average current of fuel molecules from the fuel bath to the waste bath $\langle J \rangle = L \sum_{j=1}^{m^L} (k_j^+ - \langle N_{S_j} \rangle k_j^-)$, in the case $\alpha=\beta=\gamma=0$ treated in the first part of the main text. 
Since disassembly is dominated by the spontaneous process, we expect a negligible amount of fuel molecules released by templated disassembly. Indeed, the average flux associated with a given sequence S$_j$ with error fraction $x$ reads:
\begin{align}
    L (k_j^+ - \langle N_{S_j} \rangle k_j^-) = L k_0 e^{-(\Delta \mu_r - \Delta \mu_F + ax)L} \lbk 1 - \frac{k_0 e^{-axL} + k_r e^{-\Delta \mu_F L}}{ k_0 e^{-axL} + k_r} \rbk \,,
\end{align}
where we replaced $\langle N_{S_j} \rangle$ by its expression $(k_j^+ + k_r^+)/(k_j^- + k_r^-)$. 

Since the fuel drive $\Delta \mu_F$ and the specificity $a$ are positive, the term in the bracket goes to $1$ in the large $L$ limit for $x>0$, so that $L (k_j^+ - \langle N_{S_j} \rangle k_j^-) \sim L k_j^+$.

The total flux is then given by: 
\begin{equation}
    \langle J \rangle \sim L k_0 e^{-(\Delta \mu_r - \Delta \mu_F )L} \lbk \sum_{j=1}^{m^L} e^{-a q_j} - \frac{k_0}{k_0 + k_r}\rbk\,,
\end{equation}
with $q_j$ the number of errors of sequence S$_j$. The sum is computed as follows
\begin{align}
\sum_{j=1}^{m^L} e^{-a q_j} &= \sum_{q_j=0}^{L} \Omega_{q_j} e^{-a q_j} \\
    &= \sum_{q_j=0}^{L} {L \choose q_j} ((m-1)e^{-a})^{q_j} \\
	&= \lbk 1 + (m-1) e^{-a} \rbk^L \,.
\end{align}

We recover the result from the main text, in the large $L$ limit 
\begin{equation}
    \langle J \rangle  \sim L k_0 \exp \lbk -\lp \Delta \mu_r - \Delta \mu_F - \ln \lp 1+ (m-1) e^{-a} \rp \rp L \rbk \,.
\end{equation}

\section{Choice of kinetic prefactor for the templated reaction}

In our model, the sequence-selectivity of the templated process enters via the kinetic prefactor $k_j$ involved in both the assembly and disassembly rates $k_j^+$ and $k_j^-$. 

Motivated by the experimental observation of a time delay for the incorporation of a nucleotide (monomer) following a mismatch (incorrectly copied monomer), both for DNA replication \cite{goodman_biochemical_1993} and RNA transcription \cite{thomas_transcriptional_1998}, we choose a prefactor $k_j$ that depends on the number $q$ of errors (number of incorrectly copied monomers). 
From the theoretical point of view, any function $k_j$ of the number $q$ of errors is acceptable. 

However, we show in this section that any choice other than the exponential dependence of $k_j$ on $q$ does not achieve sequence-selection. 
The reason is that kinetic rates are exponential in length $L$, since they are exponential in energies which themselves scale with $L$, and the number of sequences with error fraction $x$ is also exponential in $L$ in the large $L$ limit.

For any function $k_j$ which is super-exponential in $q=xL$, the kinetic rates of templated assembly and disassembly decay too fast, and thus both the assembly and disassembly reactions are dominated by the spontaneous process. 
The statistics of errors resulting from the spontaneous process alone is governed by combinatorial effects and dominated by random copies. This can be seen by following the derivation of \cref{app_phas_diag} but choosing $k_j = k_0 e^{-aq^b}$ with $b>1$. In this case, in the large $L$ limit, \cref{eq_ln_bw,eq_ln_fw} are modified as follows
\begin{align}
    \frac{1}{L}\ln \lp k_0 e^{-ax^b L^{b} + (- \alpha\Delta \mu_r + \gamma \Delta \mu_F )L} + k_r e^{-\beta \Delta \mu_r L} \rp = &  -\beta \Delta \mu_r + o\lp \frac{1}{L} \rp \\
    \frac{1}{L}\ln \lp k_0 e^{-ax^b L^{b} +(- \alpha\Delta \mu_r + (1+\gamma) \Delta \mu_F )L} + k_r e^{-\beta \Delta \mu_r L} \rp = & -\beta \Delta \mu_r + o\lp \frac{1}{L} \rp \,,
\end{align}
for all values of $\alpha$, $\beta$ and $\gamma$. The number of copies with error fraction $x$ is thus given by the function:
\begin{equation}
f(x) = g(x) + \frac{1}{2L} \ln \lp 2 \pi L x (1-x) \rp + \Delta \mu_r + o\lp \frac{1}{L}\rp
\end{equation}
which has a single minimum at error fraction $x_r$.

For any function $k_j$ which is sub-exponential in $q=xL$, the energetic contributions to $k_j^+$ and $k_j^-$ control the sequence-selection. 
Since 
these energies are sequence-independent in our model, the statistics of errors is governed by combinatorial effects again, and results in an average error fraction $x_r$. 
This can be seen by following the derivation of \cref{app_phas_diag} but choosing $k_j = k_0 e^{-aq^b}$ with $b<1$
In this case, in the large $L$ limit, \cref{eq_ln_bw,eq_ln_fw} are modified as follows
\begin{align}
    \frac{1}{L}\ln \lp k_0 e^{-ax^b L^b +(- \alpha\Delta \mu_r + \gamma \Delta \mu_F )L} + k_r e^{-\beta \Delta \mu_r L} \rp = & \lp - \alpha\Delta \mu_r + \gamma \Delta \mu_F  + \frac{\ln k_0-ax^b L^{b}}{L} \rp \theta \lp \gamma \Delta \mu_F - (\alpha-\beta) \Delta \mu_r \rp \nn \\
    & + \lp -\beta \Delta \mu_r + \frac{\ln k_r}{L} \rp \theta \lp  (\alpha-\beta) \Delta \mu_r - \gamma \Delta \mu_F \rp +o\lp \frac{1}{L} \rp \\
    \frac{1}{L}\ln \lp k_0 e^{-ax^b L^b +(- \alpha\Delta \mu_r + (1+\gamma) \Delta \mu_F )L} + k_r e^{-\beta \Delta \mu_r L} \rp = & \lp - \alpha\Delta \mu_r + (1+\gamma) \Delta \mu_F + \frac{\ln k_0-ax^b L^{b}}{L} \rp \theta \lp (1+\gamma) \Delta \mu_F - (\alpha-\beta) \Delta \mu_r \rp \nn \\
    & + \lp -\beta \Delta \mu_r + \frac{\ln k_r}{L} \rp \theta \lp  (\alpha-\beta) \Delta \mu_r - (1+\gamma) \Delta \mu_F \rp +o\lp \frac{1}{L} \rp 
\end{align}
Since the $x$-dependent terms in the right hand sides of the equations above decay with $L$ they vanish in the large $L$ limit. 
The number of copies with error fraction $x$ is given by the function:
\begin{align}
f(x) = & g(x) + \frac{1}{2L} \ln \lp 2 \pi L x (1-x) \rp + \Delta \mu_r \\
    & + \begin{cases}
    - \Delta \mu_F + o\lp \frac{1}{L} \rp & \text{if } (\alpha-\beta) \Delta \mu_r < \gamma \Delta \mu_F \\
    (\alpha - \beta) \Delta \mu_r - (1+\gamma) \Delta \mu_F + \frac{\ln(k_r/k_0)}{L} +ax^b L^{b-1}+ o\lp \frac{1}{L} \rp & \text{if } \gamma \Delta \mu_F < (\alpha-\beta) \Delta \mu_r < (1+\gamma) \Delta \mu_F \\
    o\lp \frac{1}{L} \rp & \text{if } (1+\gamma) \Delta \mu_F < (\alpha-\beta) \Delta \mu_r \,,
    \end{cases}
\end{align}
for all values of $\alpha$, $\beta$ and $\gamma$.
In all three cases, in the large $L$ limit the single minimum of function $f$ is $x_r$. 

To achieve sequence-selectivity, it is necessary to have an $x$-dependent correction to $f-g$ which does not vanish in the large $L$ limit, and which shifts the minimum of $f$ from $x_r$ to a lower error fraction. Since energies scale with $L$, this is only possible with an exponential dependence of $k_j$ on the number of errors $q$.

\section{Kinetic proofreading}

Kinetic proofreading introduces fuel-consuming cycles that can be completed a random number of times. Thus, when averaging over the different proofreading pathways, the coarse-grained templated process is not tightly-coupled with a well defined free energy difference between monomers and polymers anymore. As a consequence, the assembly and disassembly rates of the coarse-grained process cannot be parametrized as simply as in the absence of proofreading, and they do not obey a simple micro-reversibility condition. In the following, we show how the different proofreading pathways can be coarse-grained into a single one-step process, and we provide a simple example of proofreading dynamics.

\subsection{Explicit coarse-graining}
\label{sec_KP}

We consider the templated copy process that produces the sequence S$_j$ of length $L$ with $q$ errors described by eq. (1) in the main text. Each  proofreading pathway $\mathcal{P}$ costs an extra fuel energy $\Delta \mu_{\mathcal{P}}$ used for proofreading, in addition to $\Delta \mu_F L$ used for copying, and occurs at a rate $k_{\mathcal{P}}^+$, where we omit index $j$ for simplicity. For each proofreading pathway, a time-reversed pathway $\mathcal{P}^{\dagger}$ occurs at a rate $k_{\mathcal{P}^{\dagger}}^-$, and micro-reversibility implies that $k_{\mathcal{P}}^+/k_{\mathcal{P}^{\dagger}}^-=e^{-(\Delta \mu_r -\Delta \mu_F) L + \Delta \mu_{\mathcal{P}}}$. We choose to parametrize these rates as: 
\begin{align}
\label{eq_k+_KP}
    k_{\mathcal{P}}^+&= k_{\mathcal{P}} e^{-(\Delta \mu_r -\Delta \mu_F) L} \\
\label{eq_k-_KP}
    k_{\mathcal{P}^{\dagger}}^-&= k_{\mathcal{P}} e^{-\Delta \mu_{\mathcal{P}}}
\end{align}
with 
\begin{equation}
\label{eq_k_KP}
    k_{\mathcal{P}}=k_0^{\mathcal{P}} e^{-a(\Delta \mu_{\mathcal{P}})q} \,.
\end{equation}
Here, the specificity $a(\Delta \mu_{\mathcal{P}})$ of pathway $\mathcal{P}$ explicitly depends on the amount of energy $\Delta \mu_{\mathcal{P}}$ spent for proofreading. For any energy spent $\Delta \mu_{\mathcal{P}} \geq 0$, the specificity is increased above the level of the intrinsic specificity $a_0$ of the copy process in the absence of proofreading: $a(\Delta \mu_{\mathcal{P}}) \geq a_0$. The kinetic prefactor $k_0^{\mathcal{P}}$ is also in principle pathway-dependent. 

The rates above define kinetically-weighted probability distributions for the pathways: 
\begin{align}
\label{eq_p+_KP}
    p^{+}(\mathcal{P}) &= \frac{k_{\mathcal{P}}^{+}}{\sum_{\mathcal{P}'} k_{\mathcal{P}'}^{+}} \\ 
\label{eq_p-_KP}
    p^{-}(\mathcal{P}^{\dagger}) &= \frac{k_{\mathcal{P}^{\dagger}}^{-}}{\sum_{\mathcal{P}^{\dagger\prime}} k_{\mathcal{P}^{\dagger\prime}}^{-}} \,.
\end{align}

We now coarse-grain these different proofreading pathways into a single one-step process with an effective energy consumption and effective kinetic rates. 
The energy consumptions of the templated assembly and disassembly processes, averaged with respect to the probability distributions defined above, read $\Delta \mu_{\mathrm{en}}^+ = \langle \Delta \mu_{\mathcal{P}} \rangle_{p^+}$ and $\Delta \mu_{\mathrm{en}}^- = \langle \Delta \mu_{\mathcal{P}} \rangle_{p^-}$, where the index 'en' stands for energy. These average energies lead to the following coarse-grained chemical reaction
\begin{equation}
\label{eq_chem_reac_KP}
    \sum_{i=1}^m n_{ij} \mathrm{M}_i + \mathrm{T} + (1 + \rho^{\pm})L\mathrm{F} \ \xrightleftharpoons[k_j^-]{k_j^+} \ \mathrm{S}_j + \mathrm{T} + (1 + \rho^{\pm})L\mathrm{W} \,,
\end{equation}
where we defined $\rho^{\pm} = (\Delta \mu_{\mathrm{en}}^{\pm}/L) / \Delta \mu_F$ the per-monomer cost of proofreading, averaged over the forward ($+$) and backward ($-$) proofreading pathways, relative to the cost of copying. 

The effective coarse-grained kinetics is controlled by the average of the inverse kinetic rates \cite{gillespie_exact_1977}. The coarse-grained assembly rate $k_j^+$ and disassembly rate $k_j^-$ in \cref{eq_chem_reac_KP} are thus given by 
\begin{align}
\label{eq_inv_k+}
    k_j^+&= \frac{1}{\langle (k_{\mathcal{P}}^+)^{-1} \rangle_{p^+}} = \frac{\sum_{\mathcal{P}} k_{\mathcal{P}}^+}{N_{\mathcal{P}}}\\
\label{eq_inv_k-}
    k_j^-&= \frac{1}{\langle (k_{\mathcal{P^{\dagger}}}^-)^{-1} \rangle_{p^-} } = \frac{\sum_{\mathcal{P}{\dagger}} k_{\mathcal{P}{\dagger}}^-}{N_{\mathcal{P}}}\,,
\end{align}
where $N_{\mathcal{P}}$ is the number of proofreading pathways.
Using \cref{eq_k+_KP,eq_k-_KP,eq_k_KP}, we express these coarse-grained rates in the form of the rates considered in the main text. The assembly rate can be written as
\begin{equation}
\label{eq_kj+_kp}
    k_j^+=k_0^{\mathrm{eff}} e^{-a^+ q} e^{-(\Delta \mu_r -\Delta \mu_F) L}
\end{equation}
with the following effective kinetic prefactor $k_0^{\mathrm{eff}}$ and specificity $a^+$ in the forward dynamics:
\begin{align}
\label{eq_k0_eff}
    k_0^{\mathrm{eff}} &= \frac{\sum_{\mathcal{P}} k_0^{\mathcal{P}}}{N_{\mathcal{P}}} \\
    a^+ &= - \frac{1}{q} \ln \lbk \frac{\sum_{\mathcal{P}} k_0^{\mathcal{P}} e^{-a(\Delta \mu_{\mathcal{P}}) q}}{\sum_{\mathcal{P}} k_0^{\mathcal{P}} } \rbk \,.
\end{align}
Similarly the effective disassembly rate can be decomposed as 
\begin{equation}
\label{eq_kj-_kp}
    k_j^-=k_0^{\mathrm{eff}} e^{-a^- q} e^{-\Delta \mu_{\mathrm{kin}}}
\end{equation}
where $k_0^{\mathrm{eff}}$ is given by \cref{eq_k0_eff}, and with the following effective free energy $\Delta \mu_{\mathrm{kin}}$ (for which the index 'kin' stands for kinetic) and specificity $a^-$ in the backward dynamics:
\begin{align}
    \Delta \mu_{\mathrm{kin}} &= - \ln \lbk \frac{\sum_{\mathcal{P}}  k_0^{\mathcal{P}} e^{ - \Delta \mu_{\mathcal{P}}} }{\sum_{\mathcal{P}}  k_0^{\mathcal{P}}  } \rbk \\
    a^- &= - \frac{1}{q} \ln \lbk \frac{\sum_{\mathcal{P}}  k_0^{\mathcal{P}} e^{-a(\Delta \mu_{\mathcal{P}}) q - \Delta \mu_{\mathcal{P}}} }{\sum_{\mathcal{P}}  k_0^{\mathcal{P}} e^{ - \Delta \mu_{\mathcal{P}}} } \rbk \,.
\end{align}
%

Please note that
the forward and backward specificities $a^+$ and $a^-$ are different since in general $p^+ \neq p^-$. However, they are both larger than the intrinsic specificity $a_0$ of the copy process in the absence of proofreading, which follows from $a(\Delta \mu_{\mathcal{P}}) \geq a_0$ for all pathways $\mathcal{P}$. 
Kinetic proofreading as introduced by Hopfield \cite{hopfield_kinetic_1974} results in a squared error fraction $x_a$.
We obtain this case if $a^+= a_0 + \ln\lp 2+ e^{a_0}/(m-1)\rp$, since in the regime of accurate copying the error fraction of accurate copies is controlled by the specificity of the assembly process. In the limit of small error rates and thus large specificity, a squared error fraction is achieved by doubling the intrinsic specificity and providing an entropic correction, with $a^+ \sim 2a_0 - \ln(m-1)$.

The micro-reversibility condition is broken for the coarse-grained dynamics. Indeed, using the micro-reversibility condition for each pathway $\mathcal{P}$ we get $k_j^+/k_j^- = \sum_{\mathcal{P}} k_{\mathcal{P}}^+ / \sum_{\mathcal{P}{\dagger}} k_{\mathcal{P}{\dagger}}^- = e^{-(\Delta \mu_r -\Delta \mu_F) L} \sum_{\mathcal{P}} k_{\mathcal{P}^{\dagger}}^- e^{ \Delta \mu_{\mathcal{P}}} / \sum_{\mathcal{P}{\dagger}} k_{\mathcal{P}{\dagger}}^-$.
In the special case 
where all pathways consume the same amount of energy $\Delta \mu_{\mathcal{P}} \equiv \Delta \mu$, micro-reversibility is obeyed. In this case, it follows that $p^+ = p^-$, which implies that $\Delta \mu_{\mathrm{en}}^+=\Delta \mu_{\mathrm{en}}^-\equiv \Delta \mu$ and therefore $\rho^+=\rho^-$. It also follows that $a^+=a^- \equiv a$ and $\Delta \mu_{\mathrm{kin}} \equiv \Delta \mu$. The coarse-grained rates thus reduce to $k_j^+=k_0 e^{-a q} e^{-(\Delta \mu_r -\Delta \mu_F) L}$ and $k_j^-=k_0 e^{-a q} e^{-\Delta \mu}$, so that $k_j^+/k_j^-$ obeys the micro-reversibility condition.
In general, however, $\Delta \mu_{\mathcal{P}}$ differ. Therefore $\ln \lp \sum_{\mathcal{P}} k_{\mathcal{P}^{\dagger}}^- e^{ \Delta \mu_{\mathcal{P}}} / \sum_{\mathcal{P}{\dagger}} k_{\mathcal{P}{\dagger}}^- \rp$ differs from both average free energies $\Delta \mu_{\mathrm{en}}^+$ and $\Delta \mu_{\mathrm{en}}^-$ of forward and backward pathways and no micro-reversibility condition for the effective rates can be written.

The coarse-grained rates defined in \cref{eq_kj+_kp,eq_kj-_kp} depend on coarse grained specificities $a^\pm$ and free energy difference $\Delta\mu_{\mathrm{kin}}$.
However, the specificities $a^\pm$ could in general depend on template length $L$ and error fraction $x$, and the free energy difference $\Delta\mu_{\mathrm{kin}}$ may not scale linearly with $L$. The theory presented in the main text with fixed specificity $a$ in general constitutes the first order of the Taylor expansion of the coarse-grained parameters with respect to the parameters $L$ and $q$.
In the following section we detail a simple example where the coarse-grained parameters fulfill all assumptions.

\subsection{Single-state backtracking}

We now consider a simple proofreading model: single-state backtracking of RNA polymerase \cite{depken_intermittent_2013}. In this model, after each incorporation of a monomer to the growing copy, the polymerase can enter a backtracked state with a certain probability to correct the last added monomer. For a polymer of length $L$, the polymerase can backtrack any number of times $n$ between $0$ and $L$. There are thus $L \choose n$ proofreading pathways with $n$ backtracks, and $2^L$ proofreading pathways in total. 
Each backtrack consumes a free energy $\Delta \mu$, so the total cost of any proofreading pathway $\mathcal{P}$ with $n$ backtracks is $\Delta \mu_{\mathcal{P}}=n \Delta \mu$. We assume that all pathways have the same kinetic barrier energy, so that $k_0^{\mathcal{P}}=k_0$ for all $\mathcal{P}$. Moreover, we consider that backtracking increases the specificity linearly, as $a(n \Delta \mu)=a_0 + a_1 n/L$ with $a_0$ the intrinsic specificity. 

The probability distributions and effective parameters introduced in \cref{sec_KP} can be computed explicitly, by using the transformation $\sum_{\mathcal{P}} f(\mathcal{P}) = \sum_{n=0}^{L} {L \choose n} f(n)$ for any function $f$ that depends on a pathway $\mathcal{P}$ only via its number $n$ of backtracks. The probabilities of an assembly pathway $\mathcal{P}$ with $n$ backtracks and its time-reversed disassembly pathway $\mathcal{P}^{\dagger}$ are given by
\begin{align}
    p^+(\mathcal{P}) &= \frac{e^{-a_1 x n}}{(1+e^{-a_1 x})^L} \\
    p^-(\mathcal{P}^{\dagger}) &= \frac{e^{-(a_1 x + \Delta \mu)n}}{(1+e^{-(a_1 x+ \Delta \mu)})^L} \,,
\end{align}
where $x=q/L$ is the error fraction of sequence S$_j$.
These probabilities take the form $p^+(\mathcal{P})=(\pi^+)^n (1-\pi^+)^{L-n}$ and $p^-(\mathcal{P}^{\dagger})=(\pi^-)^n (1-\pi^-)^{L-n}$ with $\pi^+=e^{-a_1 x}/(1+e^{-a_1 x})$ and $\pi^-=e^{-a_1 x-\Delta \mu }/(1+e^{-a_1 x -\Delta \mu})$. This shows that when assembling (disassembling) the copy, the polymerase can enter the backtracked state with probability $\pi^+$ ($\pi^-$) after each monomer incorporation (removal). 

Using these expressions, it is straightforward to compute the effective free energy differences:
\begin{align}
    \Delta \mu_{\mathrm{en}}^+ &= \frac{e^{-a_1 x}}{1+e^{-a_1 x}}\Delta \mu L \\
    \Delta \mu_{\mathrm{en}}^- &= \frac{e^{-a_1 x-\Delta \mu}}{1+e^{-a_1 x-\Delta \mu}}\Delta \mu L \\
    \Delta \mu_{\mathrm{kin}} &= \ln \lp \frac{2}{1+ e^{-\Delta \mu}} \rp L \,,
\end{align}
where the first two expressions also read $\Delta \mu_{\mathrm{en}}^+ = \pi^+ \Delta \mu L$ and $\Delta \mu_{\mathrm{en}}^- = \pi^- \Delta \mu L$. The three effective energies scale with $L$.

The effective specificities read
\begin{align}
    a^+ &= a_0 + \frac{1}{x} \ln \lp \frac{2}{1+ e^{-a_1 x}} \rp\\
    a^- &= a_0 + \frac{1}{x} \ln \lp \frac{1+ e^{-\Delta \mu}}{1+ e^{-a_1 x-\Delta \mu}} \rp \,.
\end{align}
In general, the effective specifities are not independent of the error fraction $x$. They become independent of $x$ in the limit where proofreading is a small perturbation, i.e. when $a_1 \ll 1$. In this case, expanding the logarithms lead to 
\begin{align}
    a^+ &= a_0 + \frac{a_1}{2} +O(a_1^2) \\
    a^- &= a_0 +  \frac{a_1 e^{-\Delta \mu}}{1+e^{-\Delta \mu}} +O(a_1^2)\,.
\end{align}

For single-state backtracking with small specificity increment $a_1$, all the conditions listed at the end of \cref{sec_KP} are fulfilled. Therefore, this proofreading scheme can be described with the framework developed in the main text.

\section{Example: DNA replication}

DNA is a double-stranded molecule (dsDNA). Each of its strands is a polynucleotide chain composed of monomeric building blocks called deoxynucleoside monophosphate (dNMP). Those dNMP exist in four versions ($m=4$), corresponding to the four possible nucleobases that make up the nucleoside: adenine (A), cytosine (C), guanine (G) and thymine (T), and are called dAMP, dCMP, dGMP and dTMP respectively. The two strands are bound together by hydrogen bonds between facing nucleobases according to the basepairing rules: A-T and G-C. 

To be copied, the two strands must first be separated into two single-stranded DNA molecules (ssDNA) by breaking the hydrogen bonds. This operation is achieved by the enzyme helicase, which feeds on the fuel adenosine triphosphate (ATP) and releases adenosine diphosphate (ADP). The overall strand separation from start to finish reads: 
\begin{equation}
\mathrm{ds}(\mathrm{dNMP})_L + \nu L \ \mathrm{ATP} \ \xrightleftharpoons[]{} \ 2 \ \mathrm{ss}(\mathrm{dNMP})_L + \nu L \ \mathrm{ADP} \,,
\end{equation}
where $(\mathrm{dNMP})_L$ indicates the chain of dNMP of length $L$, in its double stranded or single stranded versions. The molecular motor helicase performs two tasks: moving along DNA (translocation) and breaking the hydrogen bonds, which combined consume $\nu$ molecules of ATP per basepair unwound. The value of $\nu$ may vary depending on the experimental conditions and on the helicase type, but it has been reported to have values of $\nu \approx 2$ \cite{yarranton_enzyme-catalyzed_1979}.

In order for each ssDNA to be copied, the dNMP in the environment must first be converted into deoxynucleoside triphosphate (dNTP). This reaction involves the consumption of 2 ATP molecules which become ADP molecules after losing one phosphate group to the nucleotide \cite{lieberman_enzymatic_1955}, and is catalyzed by kinases:
\begin{equation}
\mathrm{dNMP} + 2 \ \mathrm{ATP} \ \xrightleftharpoons[]{} \ \mathrm{dNTP} + 2 \ \mathrm{ADP} \,.
\end{equation}
The dNTP can then be incorporated into the growing copy of the ssDNA template. To do so, they are turned back into their monophosphate versions, which releases a pyrophosphate molecule PP$_{\rm{i}}$. The energy released from the hydrolysis of this PP$_{\rm{i}}$ is used to create the high-energy phosphodiester backbone which links the dNMP of the growing copy together. This operation is catalyzed by DNA polymerase and reads:
\begin{equation}
(\mathrm{dNMP})_n +  \mathrm{dNTP}  \ \xrightleftharpoons[]{} \ (\mathrm{dNMP})_{n+1} +  \mathrm{PP_i} \,.
\end{equation}

In our model, the different steps of polymer copying are coarse-grained into a single-step templated assembly process. In the example of DNA replication, the monomers M$_{i}$ of our model would be the dNMP, the fuel and waste molecules F and W would be the ATP and ADP molecules respectively, and the polymers S$_{j}$ would be the DNA strands $(\mathrm{dNMP})_L$.

Finally, covalent bonds in the phosphodiester backbone of DNA molecules can also be broken by spontaneous hydrolysis, resulting in two shorter DNA molecules. This reaction is slow but can be catalyzed by deoxyribonuclease enzymes (DNase) without energy expenditure. Repeated hydrolysis can break down DNA molecules into their constitutive dNMP building blocks: 
\begin{equation}
(\mathrm{dNMP})_L  \ \xrightleftharpoons[]{} \ \sum_{l=1}^{L}\mathrm{dN}_l\mathrm{MP} \,,
\end{equation}
where $\mathrm{N}_l \in \{$A,C,G,T$\}$ is the nucleobase of the nucleoside at position $l$. In our model, the repeated hydrolyses are coarse-grained into a single-step spontaneous disassembly process.


%